\documentclass[conference]{IEEEtran}

\usepackage{amssymb}
\usepackage{color,soul}
\usepackage{graphicx}
\usepackage{subcaption}
\usepackage{cite}
\usepackage{graphicx}
\usepackage{tikz}
\usepackage{subcaption}
\usepackage[cmex10]{amsmath}
\usepackage{algpseudocode}
\usepackage{algorithmicx} 
\usepackage{algorithm}
\usepackage{array}
\usepackage[hyphens]{url}
\usepackage{mathrsfs}
\usepackage{soul}
\usepackage{ulem}

\newtheorem{lemma}{Lemma}

\newtheorem{theorem}{Theorem}

\newtheorem{assumption}{Assumption}

\ifCLASSINFOpdf
\else
\fi
\begin{document}
\title{Cooperative Perception for Safe Control of Autonomous Vehicles under LiDAR Spoofing Attacks
\thanks{$*:$ Authors contributed equally. \\
This work is sponsored by NSF grant CNS-1941670.}
}
\author{\IEEEauthorblockN{Hongchao Zhang$^*$}
\IEEEauthorblockA{
Washington University \\ in St. Louis\\
hongchao@wustl.edu}
\and
\IEEEauthorblockN{Zhouchi Li$^*$}
\IEEEauthorblockA{
Worcester Polytechnic Institute \\
zli4@wpi.edu}
\and
\IEEEauthorblockN{Shiyu Cheng}
\IEEEauthorblockA{
Washington University \\ in St. Louis\\
cheng.shiyu@wustl.edu}
\and
\IEEEauthorblockN{Andrew Clark}
\IEEEauthorblockA{
Washington University \\ in St. Louis\\
andrewclark@wustl.edu}}


% \IEEEoverridecommandlockouts
% \makeatletter\def\@IEEEpubidpullup{9\baselineskip}\makeatother
% \IEEEpubid{\parbox{\columnwidth}{Permission to freely reproduce all or part
%     of this paper for noncommercial purposes is granted provided that
%     copies bear this notice and the full citation on the first
%     page. Reproduction for commercial purposes is strictly prohibited
%     without the prior written consent of the Internet Society, the
%     first-named author (for reproduction of an entire paper only), and
%     the author's employer if the paper was prepared within the scope
%     of employment.  \\
%     NDSS '16, 21-24 February 2016, San Diego, CA, USA\\
%     Copyright 2016 Internet Society, ISBN 1-891562-41-X\\
%     http://dx.doi.org/10.14722/ndss.2016.23xxx
% }
% \hspace{\columnsep}\makebox[\columnwidth]{}}
\IEEEoverridecommandlockouts
\makeatletter\def\@IEEEpubidpullup{6.5\baselineskip}\makeatother
\IEEEpubid{\parbox{\columnwidth}{
    Symposium on Vehicles Security and Privacy (VehicleSec) 2023 \\
    27 February 2023, San Diego, CA, USA \\
    ISBN 1-891562-88-6 \\
    https://dx.doi.org/10.14722/vehiclesec.2023.23066 \\
    www.ndss-symposium.org
}
\hspace{\columnsep}\makebox[\columnwidth]{}}

\maketitle

\begin{abstract}
%\boldmath
Autonomous vehicles rely on LiDAR sensors to detect obstacles such as pedestrians, other vehicles, and fixed infrastructures. LiDAR spoofing attacks have been demonstrated that either create erroneous obstacles or prevent detection of real obstacles, resulting in unsafe driving behaviors. In this paper, we propose an approach to detect and mitigate LiDAR spoofing attacks by leveraging LiDAR scan data from other neighboring vehicles. This approach exploits the fact that spoofing attacks can typically only be mounted on one vehicle at a time, and introduce additional points into the victim's scan that can be readily detected by comparison from other, non-modified scans. We develop a Fault Detection, Identification, and Isolation procedure that identifies non-existing obstacle, physical removal, and adversarial object attacks, while also estimating the actual locations of obstacles. We propose a control algorithm that guarantees that these estimated object locations are avoided. We validate our framework using a CARLA simulation study, in which we verify that our FDII algorithm correctly detects each attack pattern.
\end{abstract}

\section{Introduction}

Autonomous driving systems rely on perception modules to understand their environments, including their own positions as well as the locations of pedestrians, vehicles, and other obstacles. Perception modules develop this understanding using data from sensors such as cameras, Global Positioning Systems, RADAR, and Light Detection and Ranging (LiDAR)~\cite{wyglinski2013security}. LiDARs, which measure the distances from the LiDAR transceiver to  obstacles, provide $360^\circ$ view and 3D representation, namely point cloud, of the environment rather than a 2D image as from a camera, and thus are crucial sensors for perception in autonomous vehicles (AVs). 

Since LiDAR perception has a significant impact on the safety-critical decisions of AVs, many prior research efforts have been made to investigate the security of LiDAR perception. The LiDAR perception can be compromised by relay attacks~\cite{cao2019adversarialsen}\cite{cao2022you}\cite{shin2017illusion} and adversarial object attacks~\cite{cao2019adversarialobj}\cite{tu2020physically}. In relay attacks, a relay spoofer injects adversarial points in the point cloud, disturbs the outputs of the object detection algorithms, and either creates the perception of a fake object or hides an existing objects. In adversarial object attacks, a well-designed object  fools the object detection algorithms and makes itself undetectable. While sensor fusion based methods can partially mitigate the impact of LiDAR attacks \cite{cao2021invisible}, such methods can still be thwarted by an adversary who can target multiple sensor modalities simultaneously. 

In this paper, we propose a new approach to detect and mitigate LiDAR spoofing attacks on AVs. Our approach leverages Connected and Autonomous Vehicle~\cite{bansal2017forecasting} paradigms to share LiDAR sensor data among neighboring vehicles. Indeed, while techniques such as  Collective Perception Message (CPM) have been proposed to cooperatively perceive the environment~\cite{thandavarayan2019analysis}, to the best of our knowledge this information sharing has not been used to detect and mitigate spoofing attacks. Our approach is based on two insights. First, current demonstrated LiDAR spoofing hardwares can only target a single LiDAR sensor, and hence simultaneously spoofing multiple vehicles in a believable manner would involve a burdensome level of coordination among multiple distributed spoofers. Second, relay-based LiDAR attacks are fundamentally based on introducing new points into the LiDAR point cloud. While these spurious points may fool the targeted sensor, they will be absent from the scans of neighboring vehicles. We make the following specific contributions:
\begin{itemize}
    \item We propose a safe control system for LiDAR-perception-based AVs based on the point cloud from neighboring vehicles. In the proposed system, a Fault Detection, Identification, and Isolation (FDII) module detects and classifies the attacks, and updates the unsafe region for the vehicle. A safe controller guarantees the safety of the system based on the updated unsafe region.

    \item We analyze the correctness of the results from the FDII module. We show that the FDII module can detect and classify attacks correctly, and output the unsafe region containing the projection of the obstacles.

    \item Our results are validated through CARLA~\cite{dosovitskiy2017carla}, in which we show that the proposed FDII procedure correctly detects multiple attack types and reconstructs the true unsafe region. We then show that, under our control algorithm, the vehicle reaches the given target while avoiding an obstacle.
\end{itemize}

The rest of the paper is organized as follows. Section~\ref{sec:related-work} presents related work. Section \ref{sec:observation-and-threat-models} states the LiDAR observation model and threat model. Section \ref{sec:methodology} presents the proposed method. Section \ref{sec:case-study} contains simulation results. Section \ref{sec:conclusion} concludes the paper.
\section{Related Work}
\label{sec:related-work}
LiDAR sensors have been demonstrated to be vulnerable to spoofing attacks in \cite{shin2017illusion}. Machine learning-based  LiDAR detection was also shown to be  vulnerable in~\cite{liu2021seeing}. LiDAR spoofing attacks focus on falsifying non-existing obstacles or hiding existing obstacles. Spoofing attacks that aim to create non-existing obstacles mainly use relay attack~\cite{cao2019adversarialsen}, in which an adversary fires laser to the LiDAR measurement unit with the same wavelength to inject false points. To hide an object from being detected by LiDAR sensor, methods include adversarial objects~\cite{cao2019adversarialobj} and physical removal attacks~\cite{cao2022you}. Adversarial objects are synthesized such that deep neural network based detection modules fail to detect in a certain range of distance and angle. Physical removal attacks hide arbitrary objects, such as pedestrians, by relay attacks. 

Countermeasures to LiDAR spoofing have been proposed in recent years.  Defense approaches such as random sampling and random sampling and waveforms focus on robust perception in a single sensor scenario. Random sampling proposed in \cite{davidson2016controlling} uses robust RANSAC method to randomly sample features from the point clouds. The approach presented in \cite{matsumura2018secure} randomizes the pulses' waveforms. However, a shortcoming in practical perception of these approaches is the increase in cost. RANSAC requires high computational capability to formulate the momentum model for adversarial detection\cite{davidson2016controlling}. The approach presented in \cite{matsumura2018secure} introduces extra modulation components into the lens system and may decrease the sensitivity of LiDAR\cite{suo2020location}. A vehicle system is usually equipped with more than one sensors to estimate states and observe its surroundings. One cost-efficient method to increase robustness of estimation in faulty and adversarial environments is to use redundant information. Such a redundancy-based approach includes sensor fusion~\cite{yang2021secure} and multiple sensor overlapping. However, existing work on fault-tolerant estimation with multiple sensor overlapping focus more on the case where an agent is equipped with redundant sensors, but leave the problem of cooperative robust perception less studied.

With the development of communication and smart city, the concept of Connected Vehicles (CVs), which connect with other vehicles, pedestrians, and infrastructures, are often realized with Autonomous Vehicles (AVs) simultaneously \cite{bansal2017forecasting}. CVs provide cooperative perception that is necessary to the fully AVs which do not rely on human supervision \cite{ren2023collaborative}. The cooperative perception may also benefit the defense of LiDAR attack by providing redundant information to neighboring vehicles. However, existing research focuses on enhancing the vehicle itself by adding redundant LiDAR sensors, reducing the signal-receiving angle, transmitting pulses in random directions, and randomizing the pulses’ waveforms~\cite{shin2017illusion}\cite{pace2009detecting}\cite{pham2021survey}, and less attention has been paid on utilizing cooperative perception in the CV environment.

Recent work such as \cite{xiang2022v2xp} proposed a V2X framework that is robust to adversarial noise perturbation and collaboration attack. However, the potential of fault tolerance framework of V2X to mitigate LiDAR spoofing attacks is less studied. 

% (*Rewrite this para to review cooperative perception and VNet technologies) Cooperative perception \cite{sun2021survey} \cite{pham2021survey} utilize redundant LiDAR, randomizing the pulses’ waveforms, detector using LiDAR data in previous frames to formulate a momentum model, embedding the authentication data onto the light wave, 
\section{Observation and Threat Models}
\label{sec:observation-and-threat-models}
In this section, we introduce the LiDAR observation and threat model that we consider in this paper.

\subsection{LiDAR Observation Model}
A LiDAR sensor fires and collects $n_s$ laser beams and calculates the relative distance and angles to objects. For a laser beam indexed $i\in[1,n_s]\subseteq \mathbb{Z}^+$, we let $p_i := (s^r_i, s^a_i, s^{\phi}_i)$, where $s_i^r$ denotes the range, $s_i^a$ denotes the horizontal angle, and $s^{\phi}_i$ denotes vertical angle. A LiDAR sensor observes the environment by constructing a scan $S:=\{p_i,\ 1\leq i\leq n_s\}$. We denote the Cartesian translated LiDAR scan $S$ measured at pose $x$ as $\mathcal{O}(x,S) := \{(o_i^x, o_i^y, o_i^z), 1 \leq i \leq n_s\}$, where $o_i^x, o_i^y, \text{and }o_i^z$ denote the $x, y,$ and $z$ coordinates of $p_i$. %We let $\mathcal{P}_i$ to denote the corresponding area covered by point $p_i \in S$. %If all $n_s$ laser beams reach their furthest distances, which means the laser beam reaches the ground when $s^{\phi}_i < 0$ and the laser beam reaches the infinite when $s^{\phi}_i = 0,$ then $\cup \mathcal{P}_i, i \in [1, n_s]$ is a circle on the ground with the center at $(0, 0, 0)$ and   
% Let the scan-covered area of the observation be denoted as $\mathcal{I}$. For Agent A, denote $L:= \{j: \mathcal{I}_A \cap \mathcal{I}_j \neq \emptyset\}$ as the set of agents which have an overlapping scan-covered area with Agent A.
% Define the set of points of the obstacle as $P^{ob}.$ %Define the set of points on the surface of the obstacle as $F_{ob}.$ 
We assume that the vehicle has a default map, containing the locations of the infrastructure (for example, walls).

\subsection{Threat Model}
\label{subsec:threat-model}
% The existing LiDAR spoofing attacks are implemented by either the relay signal or the adversarial object attacks. \textcolor{red}{In this paper, we are concerned with LiDAR attacks that attempt to create fake obstacles or cancel real obstacles, and do not consider attacks on the positioning.} \textcolor{blue}{In a relay attack,} the adversary fires laser beams to inject artificial points $e^{\prime}$ into a LiDAR scan $S$. We denote the compromised LiDAR scan as $S \cup e^{\prime}$. Due to the physical limitation of the \textcolor{blue}{spoofing} hardware, the injected point can only be within a very narrow spoofing angle. %, i.e. less than $10^{\circ}$ horizontal angle. 
% Hence, in this paper, we assume that there is only one relay adversary, and the relay adversary can only spoof one LiDAR sensor. 
% \textcolor{blue}{In the adversarial object attacks, the adversarial objects are synthesized to be undetectable to the object detection algorithms of the LiDAR perception systems in a certain range of distance and angle. The adversarial objects introduce disturbance signal $e^{\prime}$ in the LiDAR scan $S$. }

% The adversarial objects, on the other hand, are synthesized to be undetectable by the LiDAR detection algorithm in a certain range of distance and angle, which may impact more than one LiDAR sensor. The attack leaves a disturbance $e^{\prime}$ at the place of the adversarial object.

\begin{figure}[!htbp]
\centering
\begin{subfigure}{.4\textwidth}
  \centering
  \includegraphics[width = \textwidth]{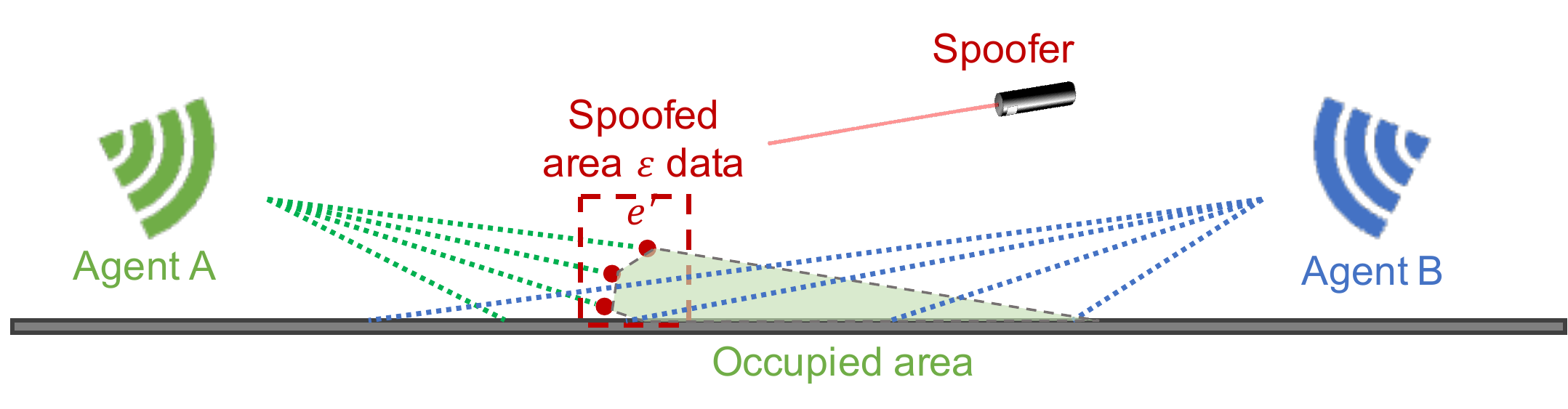}
  \subcaption{Relay attacks falsify non-existing obstacles. }
  \label{fig:RelayExp}
\end{subfigure}%
\vfill
\begin{subfigure}{.4\textwidth}
  \centering
  \includegraphics[width = \textwidth]{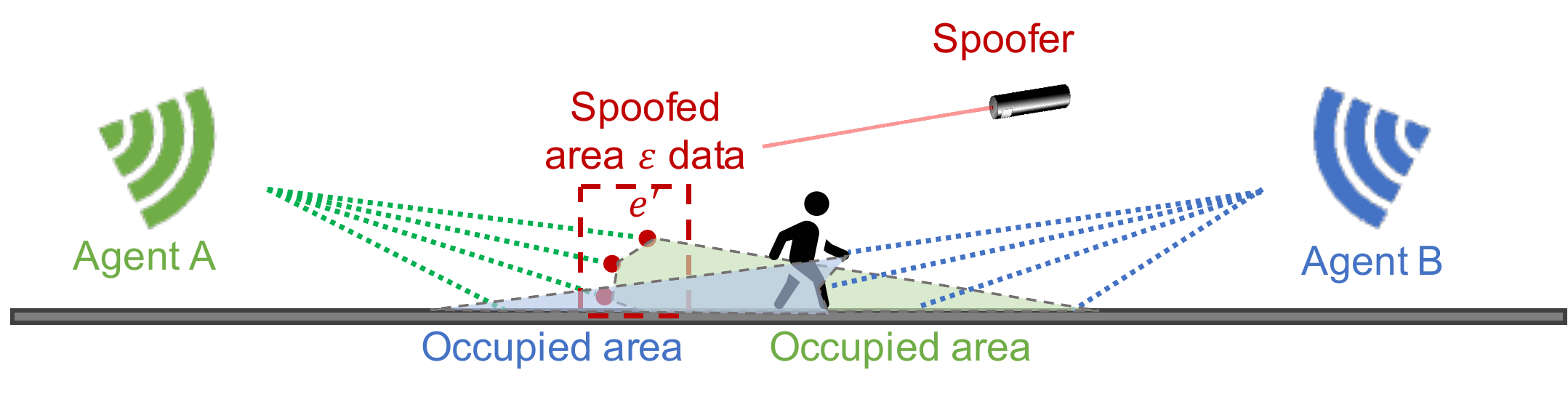}
  \subcaption{Physical removal attacks hide the object from the LiDAR detector. }
  \label{fig:PRAExp}
\end{subfigure}%
\vfill
\begin{subfigure}{.4\textwidth}
  \centering
  \includegraphics[width = \textwidth]{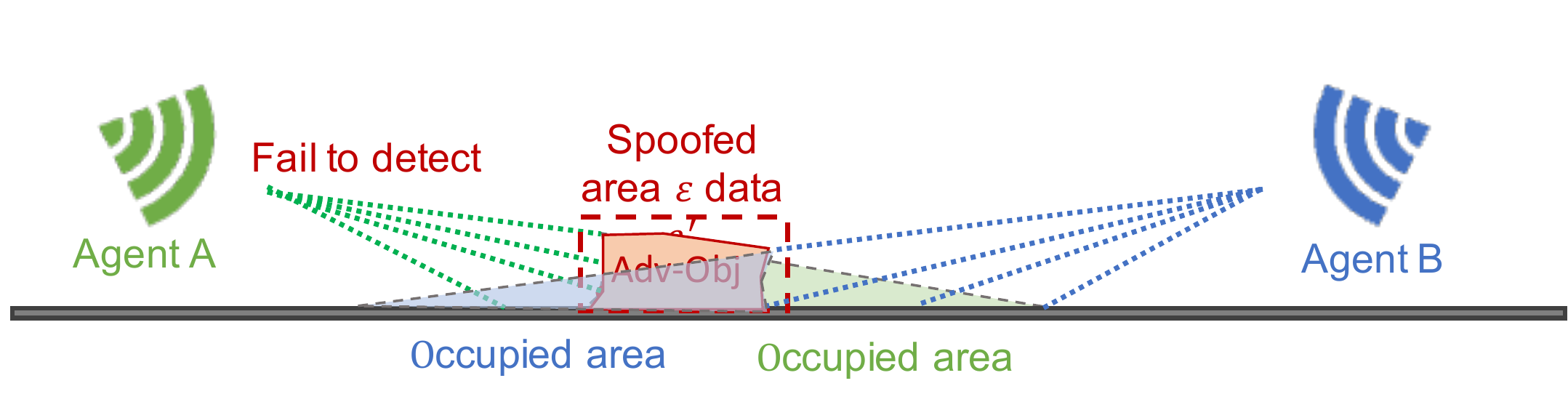}
  \subcaption{Adversarial objects can hide from the LiDAR detector. }
  \label{fig:AdvObjExp}
\end{subfigure}%
\caption{Illustration of three attack types against LiDAR-based perception. Each attack type leaves a trace in the raw data that can be detected using our proposed approach.}
\label{fig:three-LiDAR-attacks}
\end{figure}

The purpose of the attacks is to disrupt the output of object detection algorithms of the LiDAR perception by either falsifying non-existing obstacles or hiding existing obstacles. Attacks on the positioning of the vehicle are out of scope. In this paper, we consider three attacks with different purposes and implementation methods, as shown in Fig.~\ref{fig:three-LiDAR-attacks}.

\underline{\textit{Non-Existing Obstacle (NEO):}}
The spoofer falsifies non-existing obstacles by introducing relay perturbations (Fig.~\ref{fig:RelayExp}). 
In a relay attack, the adversary fires laser beams to inject artificial points $e^{\prime}$ into a LiDAR scan $S$. The resulting scan has points $S \cup e^{\prime}$. Due to the physical limitation of the spoofing hardware, the injected point can only be within a very narrow spoofing angle.  
Hence, in this paper, we assume that the relay adversary can only spoof one LiDAR sensor.
As a result of the attack, the LiDAR detection algorithm incorrectly detects an obstacle between the spoofer and the LiDAR sensor.  

% \textcolor{blue}{\underline{\textit{Non-existing Obstacle (NEO):}}}
% As shown in Fig.~\ref{fig:RelayExp}, the spoofer falsifies non-existing obstacles by \textcolor{blue}{introducing} relay perturbations. Only one LiDAR is compromised by the spoofer. The LiDAR detection algorithm incorrectly detects an obstacle between the spoofer and the LiDAR sensor. In the scan $S$ of the compromised LiDAR, there are artificial points $e^{\prime}$ corresponding to the non-existing obstacle. 

\underline{\textit{Physical Removal Attack (PRA):}}
As shown in Fig.~\ref{fig:PRAExp}, the spoofer implements the physical removal attacks by sending a relay signal to the LiDAR receiver. The LiDAR detection algorithm believes that the true obstacle does not exist. In the scan $S$ of the compromised LiDAR, there are artificial points $e^{\prime}$ between the true obstacle and the LiDAR. In order to realize the attack successfully, artificial points $e^{\prime}$ will reach the LiDAR receiver first and obscure the true obstacle. The true LiDAR signal reflected by the obstacle will be discarded by the LiDAR receiver. Due to the physical limitation of the spoofing hardware, only one LiDAR is compromised by the spoofer. 

We further divide Attack PRA into three categories, based on whether the area containing the artificial points is fully observed by the uncompromised LiDAR sensor (PRA1), partially observed (PRA2), or not observed (PRA3).

% Based on whether the uncompromised LiDAR sensor (Agent B in Fig.~\ref{fig:PRAExp}) observes the area of the artificial points $e^{\prime},$ we further divide Attack PRA into three categories.

% \underline{\textit{Attack PRA1:}}
% In this case, Agent B observes the whole area of the artificial points $e^{\prime}.$

% \underline{\textit{Attack PRA2:}}
% In this case, Agent B observes the partial area of the artificial points $e^{\prime}.$

% \underline{\textit{Attack PRA3:}}
% In this case, Agent B cannot observe the area of the artificial points $e^{\prime}.$

\underline{\textit{Adversarial Object (AO):}}
Adversarial objects are synthesized to be undetectable to the object detection algorithms of the LiDAR perception systems in a certain range of distance and angle (Fig.~\ref{fig:AdvObjExp}). The adversarial objects introduce disturbance signal $e^{\prime}$ in the LiDAR scan $S$. More than one LiDAR sensor may be compromised by one adversarial object.

In this paper, we assume that at most one attack occurs. The attack could be any of attacks NEO, PRA, or AO, and the autonomous vehicle (AV) does not know the attack type a priori.
\section{Proposed Detection and Control Approach}
\label{sec:methodology}
\begin{figure*}
    \centering
    \includegraphics[width=0.65\textwidth]{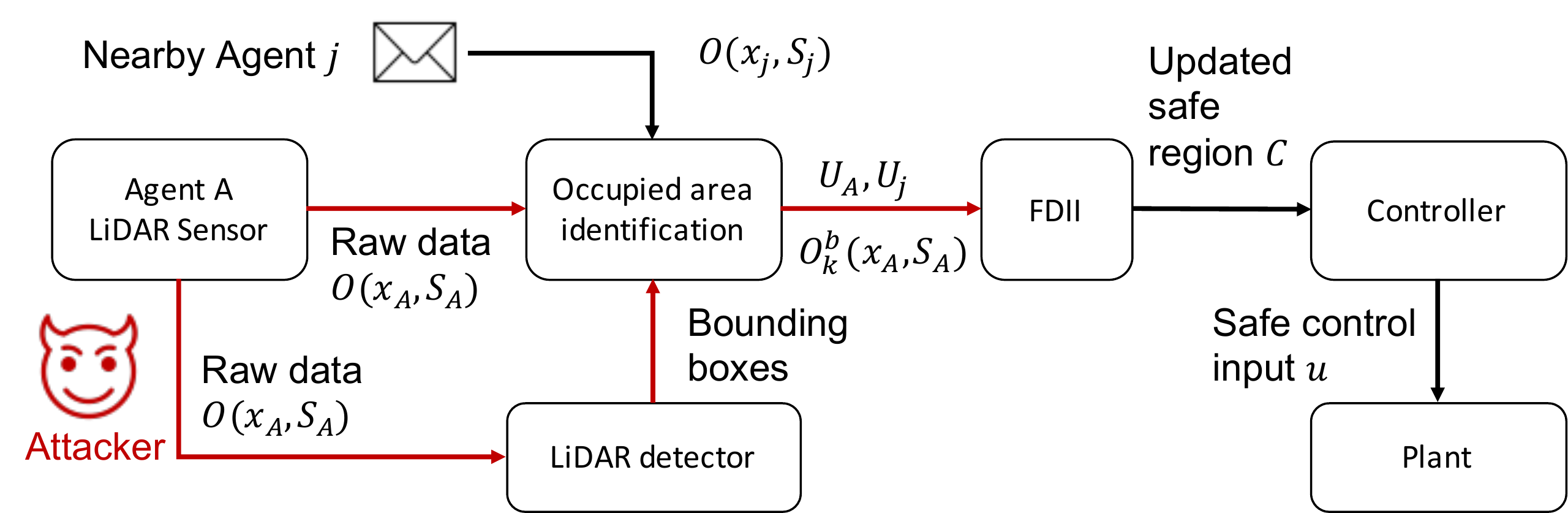}
    \caption{Schematic illustration of the proposed approach: to identify attacks marked in red, Agent A requests point cloud from nearby agents, i.e., Agent j. Then, FDII module takes $\mathcal{U}_A, \mathcal{U}_j, \mathcal{O}_k^b(x_A, S_A)$ and outputs detected attack type and updated safe region $\mathcal{C}$. Finally, controller output safe control input $u$. }
    \label{fig:illustration}
\end{figure*}

In this section, we propose a detection and control approach to ensure safety of the vehicle in an adversarial environment. The proposed approach is illustrated as Fig.~\ref{fig:illustration}. 
The victim agent leverages LiDAR observations from neighboring agents to detect and identify faults in two steps, namely, occupied area identification and the FDII, which are described as follows.

% We first describe how each vehicle detects objects from a LiDAR scan, followed by our proposed FDII approach. We then present our proposed safe control methodology that integrates the result of the FDII algorithm.

%The vehicle collects environmental perception from the neighboring vehicles, detects the attacks and updates the unsafe region based on the environmental perception, and executes safe control signal based on the updated unsafe region.
% In this section, we propose a fault-tolerant augmented perception to obtain an accurate environmental estimation. The proposed approach is illustrated as Fig.~\ref{fig:illustration}. At each time, Agent A collects the current observations of nearby agents $\mathcal{O}(x_j, S_j), \forall j \in L, j \neq A$ and calculates necessary information. Then the occupied areas are computed based on $\mathcal{O}(x_j, S_j), \forall j \in L$. Next, the FDII module detects and classifies the attacks, and updates the unsafe region based on the occupied areas.

\subsection{Occupied Area Identification}
\label{subsec:occupied-area-identification}
In this subsection, we present our method to identify and extract the area that either contains obstacles or is not observable by the LiDAR, which we denote the occupied area. The detected occupied area will then be combined with the information from other vehicles to detect attacks and compute the unsafe region (Section~\ref{sec:FDII}).

% \mathcal{I}_j = \mathcal{P}(\mathcal{O}_{max}(x,S)), where O_max denoting the pruned observation within maximum perception range.
We first prune the raw data of the observation $\mathcal{O}(x, S)$ by removing the points corresponding to the infrastructure (for example, walls) on a default map. For the rest of the paper, we use $\mathcal{O}(x, S)$ to denote the pruned observation. %The remaining points 
% We define the scan-covered area $\mathcal{I}_j = \mathcal{P}(\mathcal{O}(x, S))$ under $\mathcal{O}^b_k(x, S) = \emptyset$ as the projection of the observation of Agent $j$ if there is no obstacle. 
% Define $r_j$ as the maximum range that can be observed by Agent $j$. We define the scan-covered area $\mathcal{I}_j = \mathcal{P}(\mathcal{O}(x, S)), \mathcal{O}^b_k(x, S) = \emptyset$ as the area on the default map that can be observed by the Agent $j$ if there is no obstacle. 
% Let the true observed area of the observation be denoted as $\mathcal{I} = \mathcal{P}(\mathcal{O}(x, S))$. 
For Agent A, denote $L:= \{j: \mathcal{I}_A \cap \mathcal{I}_j \neq \emptyset\}$ as the set of agents which have an overlapping scan-covered area with Agent A. 
We define the vertical projection operation $\mathcal{P}(\mathcal{O}(x, S))\longrightarrow \mathcal{Y},$ which takes an observation $\mathcal{O}(x, S)$ as input and outputs the set $\mathcal{Y}$ of the projections of the points in $\mathcal{O}(x, S)$ onto the $x\text{-}y$ coordinate plane. %Denote the set of points corresponding to the obstacles as $\mathcal{O}^b_k(x,S) := \{p=(o^x, o^y, o^z), o^z > \zeta^z\}$. 
We define a non-obstacle scan-covered area $\mathcal{I}_j = \mathcal{P}(\bar{\mathcal{O}}(x, S))$ as the projection of the observation of Agent $j$, where $\bar{\mathcal{O}}(x, S)$ denotes the pruned observation within maximum LiDAR perception range. 

We utilize the altitude coordinates of the points to distinguish the ground and the obstacle as described in~\cite{qian2014ncc}\cite{bogoslavskyi2017efficient}. We let $\zeta^z$ denote the estimation error of the altitude coordinate.
The detection algorithm identifies $p_i = (o_i^x, o_i^y, o_i^z)$ as belonging to the ground if $o_i^z \leq \zeta^z$ and belonging to an obstacle if $o_i^z > \zeta^z.$
% For point $p_i = (o^x_i, o^y_i, o^z_i)$ in $\mathcal{O}(x, S),$ $\forall i \in n_s,$ if $o^z_i \leq \zeta^z,$ where $\zeta^z$ is the estimation error of the altitude coordinate, then $p_i$ belongs to the ground. If $o^z_i > \zeta^z,$ then $p_i$ is part of an obstacle. This scheme has been used to remove the ground from 3D scans~\cite{qian2014ncc}\cite{bogoslavskyi2017efficient}.
% \textcolor{red}{either detected by Agent $j$ or hide by Attack PRA/AO}
After ground removal, we use the 3D bounding box provided by the LiDAR-based 3D object detection algorithms of the autonomous vehicles (AVs)~\cite{zamanakos2021comprehensive} to cluster the points into a collection of obstacles, denoted $O_k^b(x,S) \subseteq O(x,S).$ Each obstacle has a corresponding bounding box denoted $\mathcal{B}_{k} \subseteq \mathbb{R}^{3}$. Here, the index $k$ ranges from $1$ to $n_o^j,$ where $n_o^j$ denotes the number of obstacles detected by Agent $j$. Note that the set $O_1^b(x,S),\ldots,O_{n_o^j}^b(x,S)$ may contain fake obstacles introduced by an adversary.
% After ground removal, we obtain points corresponding to the observations of the obstacles $\mathcal{O}^b_k(x,S) := \{p=(o^x, o^y, o^z)\textcolor{blue}{:} o^z > \zeta^z\}$, where $k \in \{1,\ldots,n^j_o\}$ is the index of the obstacles (including any non-existing obstacles introduced by Attack NEO) observed by Agent $j$ and $n^j_o$ is the number of the obstacles observed by Agent $j.$ To determine the boundaries of the observations $\mathcal{O}^b_k(x,S)$, we use the 3D bounding box provided by the LiDAR-based 3D object detection algorithms of the autonomous vehicles~\cite{zamanakos2021comprehensive}. The bounding box describes the location, size, and orientation of an object in 3D space. We consider the points that belong to one bounding box as the observation of one obstacle. 
 We regard the points that do not belong to any bounding box as the points introduced by Attack PRA or Attack AO. 
% We incorporate the index of the 
% Since there is at most one \textcolor{blue}{Attack PRA} or \textcolor{blue}{Attack AO} which hides the object from the object detection algorithms, we regard the points that do not belong to any bounding box as the points of one obstacle. %Denote the average of the points in $\mathcal{O}^o(x, S)$ as $\Bar{o}$, and the maximum distance between the points in $\mathcal{O}^o(x, S)$ and $\Bar{o}$ as $d = \max \|\|$
We further calculate the oblique projections $\mathcal{O}^p_k(x, S)$ of the points in $\mathcal{O}^b_k(x, S)$, which consists of the points where a straight line from the sensor to each point in $\mathcal{O}^b_k(x, S)$ intersects the ground. We have that $\mathcal{O}^{p}_{k}(x,S) \subseteq \mathbb{R}^{3}$. 
% We further calculate the oblique projections $\mathcal{O}^p_k(x, S)$ of the points in $\mathcal{O}^b_k(x, S)$ as follows. Suppose $p \in \mathcal{O}^b_k(x, S).$ We draw a line from the LiDAR to $p$. The line intersects the ground at point $p_g$, which is regarded as the element of $\mathcal{O}^p_k(x, S).$
% We further calculate the oblique projections of the points in $\mathcal{O}^b_k(x, S)$ from LiDAR to the ground and denote the set of the oblique projections as $\mathcal{O}^p(x, S).$ The points in $\mathcal{O}^p(x, S)$ are calculated as follows. Suppose $p \in \mathcal{O}^b_k(x, S).$ We draw a line between $p$ and the LiDAR, and choose the point on the line with $o^z = 0$ as the element of $\mathcal{O}^p(x, S).$ %These projections are calculated via similar triangles. 

We let $\mathcal{U}_{jk}$ denote the occupied area corresponding to Obstacle $k$ observed by Agent $j$, which consists of the area that contains the obstacle as well as the area that is obscured by the obstacle. %\textcolor{red}{Suppose that the proposed approach is executed on Agent A. Then Agent A calculates the occupied area $\mathcal{U}_{jk}$ from Agent $j$'s location and view. Note that Agent A uses its own bounding boxes to calculate $\mathcal{U}_{jk}.$}
Formally, the occupied area $\mathcal{U}_{jk}$ is computed as the convex hull of $\mathcal{P}(\mathcal{O}^b_k(x, S)) \cup \mathcal{O}_k^p(x, S)$. The convex hull can be calculated via open-source tools such as Scipy~\cite{virtanen2020scipy}. 
In order to compute the convex hull, we first obtain a collection of pairs of points $\{(x_1^l,y_1^l), (x_2^l, y_2^l): l=1,\ldots,n_{jk}^h\}$ from the object detection and oblique projection algorithms, where $n_{jk}^h$ is the number of the boundaries of Obstacle $k$ observed by Agent $j$. For each pair indexed $l$, we compute a half-plane constraint $h^{jk}_l(x) \leq 0$ with $h^{jk}_l(x) =  a^{jk}_l \times x + b^{jk}_l$ where $a^{jk}_l$ and $b^{jk}_l$ are defined as $a^{jk}_l = \frac{y^l_2 - y^l_1}{x^l_2 - x^l_1}$ and $b^{jk}_l = \frac{x^l_2 \times y^l_1 - x^l_1 \times y^l_2}{x^l_2 - x^l_1}.$ The convex hull is equal to the intersection of the half-plane constraints, i.e., 
$$\mathcal{U}_{jk} = \bigcap_{l=1}^{n_{jk}^h}{\{x : h_{l}^{jk}(x) \leq 0\}}.$$
% The convex hull contains all points in $\mathcal{P}(\mathcal{O}^b_k(x, S)) \cup \mathcal{O}^p(x, S)$ and the segments between those points. The boundaries $h^{jk}_l(x) =  a^{jk}_l \times x + b^{jk}_l = 0$ of the convex hull corresponding to the occupied area $\mathcal{U}_{jk}$ are a collection of line segments, where $l \in \{1,\ldots,n^h_{jk}\}$ is the number of the boundaries of obstacle $k$ observed by Agent $j$, $a^{jk}_l = \frac{y^l_2 - y^l_1}{x^l_2 - x^l_1}$, $b^{jk}_l = \frac{x^l_2 \times y^l_1 - x^l_1 \times y^l_2}{x^l_2 - x^l_1},$ and $(x^l_1, y^l_1),$ $(x^l_2, y^l_2)$ are the two vertices of the boundary. The vertices of the boundaries are the points in $\mathcal{P}(\mathcal{O}^b_k(x, S)) \cup \mathcal{O}^p(x, S)$. For $x$ that satisfies $h^{jk}_l(x) \leq 0, \forall l \in \{1,\ldots,n^h_{jk}\},$ it belongs to the convex hull. 
% The convex hull chooses some points in $\mathcal{P}(\mathcal{O}^b_k(x, S)) \cup \mathcal{O}^p(x, S)$ and uses the segments between those chosen points to surrounding all points in $\mathcal{P}(\mathcal{O}^b_k(x, S)) \cup \mathcal{O}^p(x, S)$ and the segments between all points. 
% The convex hull is calculated as the envelope lines that surround the points in $\mathcal{P}(\mathcal{O}^b_k(x, S)) \cup \mathcal{O}^p(x, S)$. The vertices of the envelope lines are the elements of $\mathcal{P}(\mathcal{O}^b_k(x, S)) \cup \mathcal{O}^p(x, S)$. 

The occupied area computed by the above procedure may not fully contain the obstacle due to the fact that the obstacle location must be interpolated from a finite number of samples. We enhance the robustness of these sampling errors by changing the boundaries of the occupied area to $h_l^{jk}(x) - ||a_{l}^{jk}||\zeta_{h} \leq 0,$ where $\zeta_h = \zeta_n + \zeta_r$, where $\zeta_n$ is the observation noise bound and $\zeta_r$ is a bound on the distance between neighboring sample points that can be obtained from the distance to the object and the angular resolution of the LiDAR. We assume that the LiDAR resolution is sufficiently large such that each point on the obstacle that is in the line-of-sight of the LiDAR is at most $\zeta_r$ distance away from at least one scan point.

In what follows, we will show that each obstacle visible to Agent $j$ (including false or adversarial objects) is contained in an occupied region that is computed according to the procedure described above. Let $P^{ob} \subseteq \mathbb{R}^{3}$ denote the location of an obstacle.
% For each obstacle, define the set of points of the obstacle as $P^{ob}.$ 
We divide the obstacle into two parts. The first part is the set of points in $P^{ob}$ that have line-of-sight with the LiDAR. We denote the first part as $P_1.$ The second part, which is denoted as $P_2,$ is the set of points in $P^{ob}$ that do not have line-of-sight with the LiDAR.

\begin{assumption}
    \label{assmpt:no-obscure}
    For Agent $j \in L$ with occupied areas $\mathcal{U}_{jk}, k \in \{1,\ldots,n^j_o\}$ and Obstacle $k^\prime \in \{1,\ldots,n_o^j\}$ with the set of points $P^{ob}_{k^{\prime}} \subseteq \mathbb{R}^{3},$ we have $\mathcal{P}(P^{ob}) \bigcap (\cup_{k \in \{1,\ldots,n^j_o\} \setminus \{k^\prime\}} \mathcal{U}_{jk}) = \emptyset.$
    % \textcolor{red}{For each Obstacle $k,$ its $P_1$ can be observed by an Agent $j, j \in L,$ which is equivalent to $P_1 \subseteq \mathcal{O}^b_k(x_j,S_j).$}
\end{assumption}

Intuitively, Assumption \ref{assmpt:no-obscure} implies that for any obstacle visible to agent j, there is no overlap between the obstacle and the occupied area of any other obstacle detected by agent j. 

\begin{lemma}
\label{lemma:occupied-area-covers-obstacle-projection}
    Suppose that  Assumption~\ref{assmpt:no-obscure} holds and we are given the observation of an obstacle $\mathcal{O}^b_k(x,S)$ in a bounding box and the set of oblique projections $\mathcal{O}^p(x, S).$ If occupied area $\mathcal{U}$ is computed as the convex hull of $\mathcal{P}(\mathcal{O}^b_k(x, S)) \cup \mathcal{O}^p(x, S)$, then $\mathcal{P}(P^{ob}) \subseteq \mathcal{U}.$%then the vertical projections of the points of the obstacle are in $\mathcal{U}$.
\end{lemma}

\begin{proof}
    As described above, $P^{ob} = P_1 \cup P_2.$ In what follows, we describe how $\mathcal{P}(P_1) \subseteq \mathcal{U}$ and $\mathcal{P}(P_2) \subseteq \mathcal{U},$ and hence $\mathcal{P}(P^{ob}) \subseteq \mathcal{U}.$
    % The obstacle $P^{ob}$ consists of two parts: the part $P_1$ that can be observed by a certain agent (eg. Agent A) with observation $\mathcal{O}^b_k(x, S),$ and the part $P_2$ that cannot be observed by Agent A (the interior and back side corresponding to Agent A's position).

    \underline{\textit{$P_1$}}: 
    Let $x \in P_{1}$. By assumption, since every point in $P_{1}$ is in the line of sight of the LiDAR, there exists a sample point $x^{\prime}$ such that $||x^{\prime}-x|| \leq \zeta_{h}$. Hence, for all $l=1,\ldots,n_{jk}^{h}$, we have 
    \begin{eqnarray*}
    h_{l}^{jk}(x) &=& a_{l}^{jk}x + b_{l}^{jk} \\
    &=& a_{l}^{jk}(x-x^{\prime}) + a_{l}^{jk}x^{\prime} + b_{l}^{jk} \\
    &\leq& ||a_{l}^{jk}||\zeta_{h} + b_{l}^{jk} + a_{l}^{jk}x^{\prime} \\
    &\leq& ||a_{l}^{jk}||\zeta_{h} 
    \end{eqnarray*}
   where the first inequality follows from Cauchy-Schwartz and the second inequality follows from the fact that $h_{l}^{jk}(x^{\prime}) \leq 0$ for all scan points $x^{\prime}$. Hence $x$ lies in the occupied area.

  %  By assumption, since every point in $P_1$ is in the line-of-sight of the LiDAR, it is at most $\zeta_r$ distance from at least one sample point. Hence, for all $l=1,\ldots,n_{jk}^{h}$, we have $$h_{l}^{jk}(x) =  \leq \|a_l^{jk} \times \zeta_h + b_l^{jk}\|$$
    % The points in $P_1$ either belong to $\mathcal{O}^b_k(x, S)$ or are between two neighboring scan points. 
    %For the points that belong to $\mathcal{O}^b_k(x, S)$, according to the calculation of the convex hull, the projections of the points that belong to $\mathcal{O}^b_k(x, S)$ are contained in the convex hull.
    % For the points that belong to $\mathcal{O}^b_k(x, S)$, since the convex hull surrounds the points in $\mathcal{P}(\mathcal{O}^b_k(x, S)) \cup \mathcal{O}^p(x, S)$, the projections of the points that belong to $\mathcal{O}^b_k(x, S)$ are contained in the convex hull. 
   % For the points that are between two neighboring scan points, since $\max_x h_l^{jk}(x) \leq \|a_l^{jk} \times \zeta_h + b_l^{jk}\|,$ they are contained in the convex hull $h_l(x) - \|a_l^{jk} \times \zeta_h + b_l^{jk}\| \leq 0.$
    % For the points that are between two neighboring scan points, they are contained in the convex hull guaranteed by the margin $\zeta_h.$

    \underline{\textit{$P_2$}}: Suppose there is a point $p = (s^r, s^a, s^{\phi}) = (o^x, o^y, o^z) \in P_2.$ There must exist a point $p_{\ast} = (s^r_{\ast}, s^a_{\ast}, s^{\phi}_{\ast}) = (o^x_{\ast}, o^y_{\ast}, o^z_{\ast}) \in P_1$ such that $s^a = s^a_{\ast}$, $s^{\phi} = s^{\phi}_{\ast}$ and $s^r > s^r_{\ast},$ which block the line-of-sight between the LiDAR and $p.$ We denote the oblique projection of $p$ and $p_{\ast}$ as $p_o$. Since $s^a = s^a_{\ast}$ and $s^{\phi} = s^{\phi}_{\ast}$, the oblique projections of $p$ and $p_{\ast}$ are identical, which means that $p$, $p_{\ast}$, and $p_o$ are collinear. Hence, the projection $\mathcal{P}(p)$, the projection $\mathcal{P}(p_\ast)$, and $\mathcal{P}(p_o)$ are colinear.

   % Next we would like to show that the projection $\mathcal{P}(p)$, the projection $\mathcal{P}(p_{\ast})$, and $p_0$ are collinear. 
    %We have $\overrightarrow{p_0p} = (o^x, o^y, o^z)$ and $\overrightarrow{p_0p_{\ast}} = (o^x_{\ast}, o^y_{\ast}, o^z_{\ast})$. Since $p$, $p_{\ast}$, and $p_0$ are collinear, we have $\overrightarrow{p_0p} \times \overrightarrow{p_0p_{\ast}} = 0,$ which gives us
    %\begin{align}
     %   \label{eq:collinear-obstacle-1}
     %   o^yo^z_{\ast} - o^zo^y_{\ast} &= 0 \\
     %   \label{eq:collinear-obstacle-2}
     %   o^zo^x_{\ast} - o^xo^z_{\ast} &= 0 \\
     %   \label{eq:collinear-obstacle-3}
     %   o^xo^y_{\ast} - o^yo^x_{\ast} &= 0 
    %\end{align}
    %Similarly, we have $\overrightarrow{p_0\mathcal{P}(p)} = (o^x, o^y, 0)$ and $\overrightarrow{p_0\mathcal{P}(p_{\ast})} = (o^x_{\ast}, o^y_{\ast}, 0)$. Then we can get $\overrightarrow{p_0\mathcal{P}(p)} \times \overrightarrow{p_0\mathcal{P}(p_{\ast})} = (0, 0, o^xo^y_{\ast} - o^yo^x_{\ast})$. According to Eq.~\eqref{eq:collinear-obstacle-3}, we have $\overrightarrow{p_0\mathcal{P}(p)} \times \overrightarrow{p_0\mathcal{P}(p_{\ast})} = (0, 0, 0),$ which means that the projection $\mathcal{P}(p)$, the projection $\mathcal{P}(p_{\ast})$, and $p_0$ are collinear.

    Finally, we  show that $\mathcal{P}(p)$ belongs to the convex hull. Since the convex hull contains $P_1$ and its oblique projection, $\mathcal{P}(p_{\ast})$ and $p_0$ belong to the convex hull. Since the convex hull is a convex set, it contains the line segment $\overline{p_0\mathcal{P}(p_{\ast})}.$ Since the projection $\mathcal{P}(p)$, the projection $\mathcal{P}(p_{\ast})$, and $p_0$ are collinear, $\mathcal{P}(p)$ belongs to the line segment $\overline{p_0\mathcal{P}(p_{\ast})},$ which means that $\mathcal{P}(p)$ belongs to the convex hull.

    Since both the projections of the points in $P_1$ and $P_2$ belong to the convex hull, and the occupied area $\mathcal{U}$ is computed as the convex hull of $\mathcal{P}(\mathcal{O}^b_k(x, S)) \cup \mathcal{O}^p(x, S)$, thus we have $\mathcal{P}(P^{ob}) \subseteq \mathcal{U}.$ 
    % \underline{\textit{$P_2$}}: Suppose there is a point $p = (s^r, s^a, s^{\phi}) = (o^x, o^y, o^z) \in P_2.$ There must exist a point $p_{\ast} = (s^r_{\ast}, s^a_{\ast}, s^{\phi}_{\ast}) = (o^x_{\ast}, o^y_{\ast}, o^z_{\ast}) \in P_1$ such that $s^a = s^a_{\ast}$,  and $s^{\phi} < s^{\phi}_{\ast}$
\end{proof}

\subsection{Fault Detection, Identification, and Isolation}
\label{sec:FDII}
\begin{figure}
    \centering
    \includegraphics[width=0.3\textwidth]{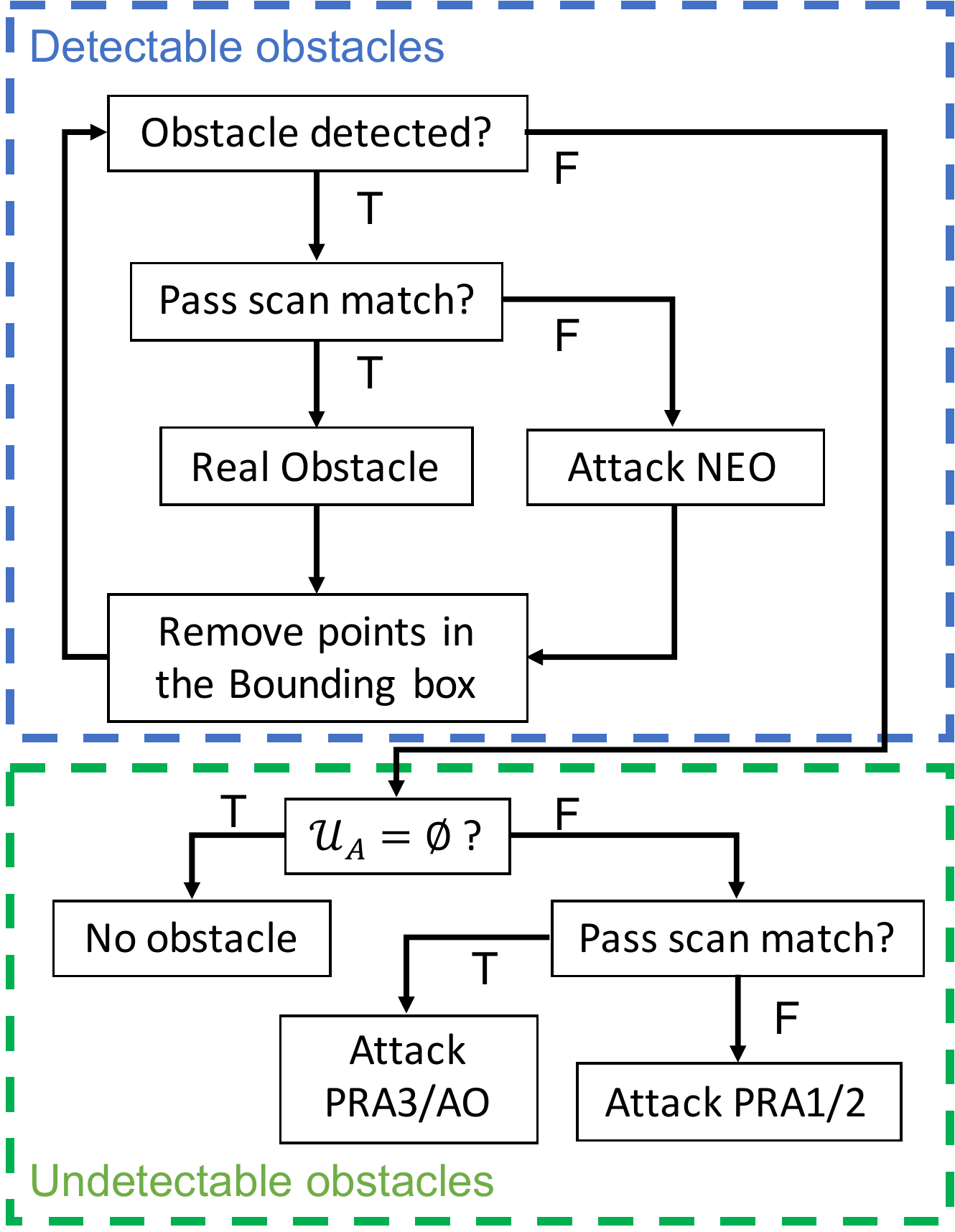}
    \caption{Decision tree of the FDII module: in the blue box, we iterate over detectable obstacles to detect faults. Then we remove points contained in bounding boxes and pass the remaining point cloud to the green box to identify undetectable obstacles. }
    \label{fig:FDII}
\end{figure}

The objective of the proposed fault detection, identification, and isolation (FDII) module is to detect the deviations between the observations of different agents, identify the spoofed agent, isolate the corrupted parts of the raw data, and restore the true unsafe region. The proposed FDII module is illustrated as Fig.~\ref{fig:FDII}. FDII first iterates over all detected obstacles to detect faults by leveraging LiDAR scan data from other neighboring vehicles.  It then removes points contained in the bounding box and uses the residual point cloud for  false data detection and identification.

% \begin{table}[]
%     \centering
%     \begin{tabular}{ |c|c|c|c| } 
%      \hline
%      Type of Attack & detect i & $U_i$ Non-Empty & Scan Match \\ 
%      \hline
%      None True & T & T & T \\ 
%      None False & F & F & - \\ 
%      a & T & T & F \\ 
%      b1, b2 & F & T & F \\ 
%      b3, c & F & T & T \\
%      \hline
%     \end{tabular}
%     \caption{Decision Table}
%     \label{tab:my_label}
% \end{table}

% If it is a 2-D LiDAR, we use $\Tilde{\mathcal{O}}_i^{\mathcal{I}}$ and $\Tilde{\mathcal{O}}_j^{\mathcal{I}}$. If it is a 3-D LiDAR, we use the values of x and y coordinates of $\Tilde{\mathcal{O}}_i^{\mathcal{I}}$ and $\Tilde{\mathcal{O}}_j^{\mathcal{I}}$, which means the projection of the scan on the horizontal plane. 

% Since the proposed method is based on the extra LiDAR observation, We  assume that $\mathcal{P}(\mathcal{O}^b_k(x_A, S_A)) \cap (I_B \setminus \cup(\mathcal{U}_B \setminus \mathcal{P}(\mathcal{O}^b_k(x_A, S_B)))) \neq \emptyset,$ which means that Agent B can observe at least a part of the area of the obstacle which Agent A observes. 
% Since the proposed method is based on the extra LiDAR observation, we limit the range of fault detection as $\mathcal{I}_A \cap \mathcal{I}_B.$ We further assume that the 
% The following lemma guarantees the detection of the attack if the spoofer presents.
% The following lemma guarantees the detection of the mismatch between observations even if the spoofer presents.

We first describe how Agent A incorporates LiDAR information from other neighboring agents. At each time, Agent A collects the current observations of nearby agents $\mathcal{O}(x_j, S_j), \forall j \in L, j \neq A$, position $x_j$, and sensor information including observation noise bound and resolution bound. The corresponding observations $\mathcal{O}(x_A, S_j),$ the observations of obstacles $\mathcal{O}^b_k(x_A, S_j),$ and the occupied areas $\mathcal{U}_{jk}$ are calculated by Agent A by repeating the steps described in Section~\ref{subsec:occupied-area-identification}. In $\mathcal{O}(x_A, S_j),$ Agent A marks the points in its bounding boxes as $\mathcal{O}_k^b(x_A,S_j) = \mathcal{O}(x_A,S_j) \cap \mathcal{B}_k.$ For each $\mathcal{O}^b_k(x_A, S_j),$ Agent A calculates the corresponding $\mathcal{O}^p_k(x, S).$ Agent A then calculates $\mathcal{U}_{jk}$ by computing the convex hull of $\mathcal{P}(\mathcal{O}^b_k(x, S)) \cup \mathcal{O}^p_k(x, S).$  %\textcolor{red}{After each $\mathcal{U}_{jk}$ is calculated, Agent A removes the points in the corresponding bounding box from the observation $\mathcal{O}(x_A, S_j).$}

We define the scan points in observation $\mathcal{O}(x_A, S_A)$ that is not detected by Agent A as $\mathcal{O}^u_A(x_A, S_A) := \mathcal{O}(x_A, S_A)\backslash \bigcup_k \mathcal{O}^b_k(x_A, S_A)$ and the undetected points in observation $\mathcal{O}(x_A, S_j)$ as $\mathcal{O}^u_j(x_A, S_A) := \mathcal{O}(x_A, S_j)\backslash \bigcup_k \mathcal{O}^b_k(x_A, S_j)$. We then compute their corresponding occupied area $\mathcal{U}^u_{A}$ and $\mathcal{U}^u_{j}$ by calculating their convex hulls. 

We now analyze how each of the attacks defined in Section~\ref{subsec:threat-model} affects the LiDAR perception and occupied area identification introduced in Section~\ref{subsec:occupied-area-identification}. In what follows, we use $A$ to denote the index of the victim agent and analyze whether the obstacle observed by Agent A could also be observed and validated by other agents, i.e. B under each of the attacks defined in Section~\ref{subsec:threat-model}. 

\underline{\textit{For Attack NEO}}, the artificial points $e^{\prime}$ introduce a fake obstacle with points $\mathcal{O}^b_k(x_A, S_A\cup e^{\prime})$. %Since there is only one victim, we have $\mathcal{O}^b_k(x_A, S_A) \nsubseteq \mathcal{U}_B$. 
Since there is no true obstacle, we have $\mathcal{U}_{Bk} = \emptyset$, and as a corollary, $\mathcal{P}(\mathcal{O}^b_k(x_A, S_A)) \nsubseteq \mathcal{U}_{Bk}.$

\underline{\textit{For Attack PRA1}}, the spoofed obstacle location does not overlap with the occupied area of agent B. Hence $\mathcal{P}(\mathcal{O}_{A}^{u}(x_{A}, S_{A} \cup e^{\prime})) \nsubseteq \mathcal{U}^{u}_{B}$ and $\mathcal{P}(\mathcal{O}^{u}_{A}(x_{A},S_{A} \cup e^{\prime})) \cap \mathcal{U}^{u}_{B} = \emptyset$.

% \underline{\textit{For Attack PRA1}}, the artificial points $e^{\prime}$ introduce false data into $\mathcal{O}^u_A(x_A, S_A\cup e^\prime)$. \textcolor{red}{Here $\mathcal{U}_{Bk}$ is calculated with the points that do not belong to any bounding box.} Since there is a true obstacle, $\mathcal{U}_{Bk} \neq \emptyset.$ In this case, since Agent B could observe the entire area affected by false data $e^{\prime}$, we have $\mathcal{P}(\mathcal{O}^u_A(x_A, S_A\cup e^\prime)) \nsubseteq \mathcal{U}_{Bk}$ and $\mathcal{P}(\mathcal{O}^u_A(x_A, S_A\cup e^\prime)) \cap \mathcal{U}_{Bk} = \emptyset. $ 

\underline{\textit{For Attack PRA2}}, there are also fake $\mathcal{O}^u_A(x_A, S_A\cup e^{\prime})$ and non-empty $\mathcal{U}^{u}_{B}$. However, $\mathcal{P}(\mathcal{O}^u_A(x_A, S_A\cup e^{\prime})) \nsubseteq \mathcal{U}^{u}_{B}$ and $\mathcal{P}(\mathcal{O}^u_A(x_A, S_A)) \cap \mathcal{U}^{u}_{B} \neq \emptyset$ in this case because Agent B could only observe the partial space of $\mathcal{O}^u_A(x_A, S_A\cup e^{\prime})$. 

\underline{\textit{For Attack PRA3 and AO}}, $\mathcal{P}(\mathcal{O}_{A}^{u}(x_{A}, S_{A} \cup e^{\prime}))$ is fully covered by non-empty $\mathcal{U}^{u}_{B}$, which means $\mathcal{P}(\mathcal{O}_{A}^{u}(x_{A}, S_{A} \cup e^{\prime})) \subseteq \mathcal{U}^{u}_{B}$. 

% \underline{\textit{For Attack PRA3}}, the fake $\mathcal{P}(\mathcal{O}_{A}^{u}(x_{A}, S_{A} \cup e^{\prime}))$ is fully covered by non-empty $\mathcal{U}^{u}_{B}$ because Agent B could not observe the space of $\mathcal{O}_{A}^{u}(x_{A}, S_{A} \cup e^{\prime})$, which means $\mathcal{P}(\mathcal{O}_{A}^{u}(x_{A}, S_{A} \cup e^{\prime})) \subseteq \mathcal{U}^{u}_{B}$. 

% \underline{\textit{For Attack AO}}, there is non-empty $\mathcal{O}_{A}^{u}(x_{A}, S_{A} \cup e^{\prime})$ and non-empty $\mathcal{U}^{u}_{B}$ corresponding to the adversarial object. According to Lemma~\ref{lemma:occupied-area-covers-obstacle-projection}, we have $\mathcal{P}(P^{ob}) \subseteq \mathcal{U}^{u}_{B}.$ Since $\mathcal{P}(\mathcal{O}_{A}^{u}(x_{A}, S_{A} \cup e^{\prime})) \subseteq \mathcal{P}(P^{ob})$, we have $\mathcal{P}(\mathcal{O}_{A}^{u}(x_{A}, S_{A} \cup e^{\prime})) \subseteq \mathcal{U}^{u}_{B}$.

% \underline{\textit{For Attack AO}}, there is non-empty $\mathcal{O}_{A}^{u}(x_{A}, S_{A} \cup e^{\prime})$ and non-empty $\mathcal{U}^{u}_{B}$ corresponding to the adversarial object and the true obstacle, respectively. According to Lemma~\ref{lemma:occupied-area-covers-obstacle-projection}, we have $\mathcal{P}(P^{ob}) \subseteq \mathcal{U}^{u}_{B}.$ Since $\mathcal{P}(\mathcal{O}_{A}^{u}(x_{A}, S_{A} \cup e^{\prime})) \subseteq \mathcal{P}(P^{ob})$, we have $\mathcal{P}(\mathcal{O}_{A}^{u}(x_{A}, S_{A} \cup e^{\prime})) \subseteq \mathcal{U}^{u}_{B}$.

\underline{\textit{For the true obstacle}}, there is non-empty $\mathcal{O}^b_k(x_A, S_A\cup e^{\prime})$ and non-empty $\mathcal{U}_{Bk}$ corresponding to the true obstacle. According to Lemma~\ref{lemma:occupied-area-covers-obstacle-projection}, we have $\mathcal{P}(P^{ob}) \subseteq \mathcal{U}_{Bk}.$ Since $\mathcal{O}^b_k(x_A, S_A\cup e^{\prime}) \subseteq \mathcal{P}(P^{ob})$, we have $\mathcal{O}^b_k(x_A, S_A\cup e^{\prime}) \subseteq \mathcal{U}_{Bk}$.

% The intuition of the FDII is that if the LiDAR observation of Agent A is spoofed by the adversary, the artificial points $e^{\prime}$ will not match the data collected by another agent $B \in L \setminus \{A\}$. The set $\mathcal{P}(\mathcal{O}^b_k(x_A, S_A)) \setminus \mathcal{U}_{Bk}$ is equal to the set of points that are observed by Agent A and are in the space that can be observed by Agent B. This set can be used to detect whether Obstacle $k$ is a true obstacle or an attack is taking place, as described in the following lemma.

The following lemma uses the preceding analysis to describe how the scan data from agent B can be used to detect and identify the attack type.

\begin{lemma}
    \label{lemma:scan-match}
    % Suppose Agent A observes obstacle with observation $\mathcal{O}^b_k(x_A, S_A)$ and receives $\mathcal{U}_B$ from Agent B. 
    If $\mathcal{P}(\mathcal{O}^b_k(x_A, S_A \cup e^{\prime})) \setminus \mathcal{U}_{Bk} \neq \emptyset,$ then Agent A is being targeted by an attack of type NEO. If $\mathcal{P}(\mathcal{O}^b_k(x_A, S_A\cup e^{\prime})) \setminus \mathcal{U}_{Bk} = \emptyset,$ then Obstacle $k$ is a true obstacle. 
    % If $\mathcal{P}(\mathcal{O}^b_k(x_A, S_A)) \setminus \mathcal{U}_{Bk} \neq \emptyset,$ then Agent A is being targeted by an attack of type NEO, PRA1, or PRA2. If $\mathcal{P}(\mathcal{O}^b_k(x_A, S_A)) \setminus \mathcal{U}_{Bk} = \emptyset,$ then either Obstacle $k$ is a true obstacle, or Agent A is under Attack PRA3 or AO.
If $\mathcal{P}(\mathcal{O}_{A}^{u}(x_{A}, S_{A} \cup e^{\prime})) \setminus \mathcal{U}^{u}_{B} \neq \emptyset,$ then Agent A is being targeted by PRA1 or PRA2. If $\mathcal{P}(\mathcal{O}_{A}^{u}(x_{A}, S_{A} \cup e^{\prime})) \setminus \mathcal{U}^{u}_{B} = \emptyset,$ then Agent A is under Attack PRA3 or AO.
\end{lemma}

\begin{proof}
    We first consider the detectable obstacles $\mathcal{O}^b_k(x_A, S_A\cup e^{\prime}).$ If $\mathcal{P}(\mathcal{O}^b_k(x_A, S_A\cup e^{\prime})) \setminus \mathcal{U}_{Bk} \neq \emptyset,$ which is equivalent to $\mathcal{P}(\mathcal{O}^b_k(x_A, S_A\cup e^{\prime})) \neq (\mathcal{P}(\mathcal{O}^b_k(x_A, S_A\cup e^{\prime})) \cap \mathcal{U}_{Bk}),$ then we have $\mathcal{P}(\mathcal{O}^b_k(x_A, S_A\cup e^{\prime})) \nsubseteq \mathcal{U}_{Bk}.$ According to the previous analysis of the impact of the attacks, Attack NEO occurs. If $\mathcal{P}(\mathcal{O}^b_k(x_A, S_A\cup e^{\prime})) \setminus \mathcal{U}_{Bk} = \emptyset$ and $e^{\prime} = \emptyset,$ which is equivalent to $\mathcal{P}(\mathcal{O}^b_k(x_A, S_A\cup e^{\prime})) = (\mathcal{P}(\mathcal{O}^b_k(x_A, S_A\cup e^{\prime})) \cap \mathcal{U}_{Bk}),$ then we have $\mathcal{P}(\mathcal{O}^b_k(x_A, S_A\cup e^{\prime})) \subseteq \mathcal{U}_{Bk}.$
    According to the previous analysis of the impact of the attacks, there is a true obstacle.

    We next consider the undetectable obstacles $\mathcal{O}_{A}^{u}(x_{A}, S_{A} \cup e^{\prime}).$ If $\mathcal{P}(\mathcal{O}_{A}^{u}(x_{A}, S_{A} \cup e^{\prime})) \setminus \mathcal{U}^{u}_{B} \neq \emptyset,$ which is equivalent to $\mathcal{P}(\mathcal{O}_{A}^{u}(x_{A}, S_{A} \cup e^{\prime})) \neq (\mathcal{P}(\mathcal{O}_{A}^{u}(x_{A}, S_{A} \cup e^{\prime})) \cap \mathcal{U}^{u}_{B}),$ then we have $\mathcal{P}(\mathcal{O}_{A}^{u}(x_{A}, S_{A} \cup e^{\prime})) \nsubseteq \mathcal{U}^{u}_{B}.$ According to the previous analysis of the impact of the attacks, Attack PRA1 or PRA2 occurs. If $\mathcal{P}(\mathcal{O}_{A}^{u}(x_{A}, S_{A} \cup e^{\prime})) \setminus \mathcal{U}^{u}_{B} = \emptyset,$ which is equivalent to $\mathcal{P}(\mathcal{O}_{A}^{u}(x_{A}, S_{A} \cup e^{\prime})) = (\mathcal{P}(\mathcal{O}_{A}^{u}(x_{A}, S_{A} \cup e^{\prime})) \cap \mathcal{U}^{u}_{B}),$ then we have $\mathcal{P}(\mathcal{O}_{A}^{u}(x_{A}, S_{A} \cup e^{\prime})) \subseteq \mathcal{U}^{u}_{B}.$
    According to the previous analysis of the impact of the attacks, Attack PRA3 or AO occurs. 
    % We first consider $\mathcal{P}(\mathcal{O}^b_k(x_A, S_A)) \setminus \mathcal{U}_{Bk} \neq \emptyset,$ which is equivalent to $\mathcal{P}(\mathcal{O}^b_k(x_A, S_A)) \neq (\mathcal{P}(\mathcal{O}^b_k(x_A, S_A)) \cap \mathcal{U}_{Bk}).$ Thus, we have $\mathcal{P}(\mathcal{O}^b_k(x_A, S_A)) \nsubseteq \mathcal{U}_{Bk}.$
    % According to the previous analysis of the impact of the attacks, if $\mathcal{U}_{Bk} = \emptyset$, Attack NEO occurs. If $\mathcal{U}_{Bk} \neq \emptyset$, Attack PRA1 or PRA2 occurs.
    % We next consider $\mathcal{P}(\mathcal{O}^b_k(x_A, S_A)) \setminus \mathcal{U}_{Bk} = \emptyset,$ which is equivalent to $\mathcal{P}(\mathcal{O}^b_k(x_A, S_A)) = (\mathcal{P}(\mathcal{O}^b_k(x_A, S_A)) \cap \mathcal{U}_{Bk}).$ Thus, we have $\mathcal{P}(\mathcal{O}^b_k(x_A, S_A)) \subseteq \mathcal{U}_{Bk}.$
    % According to the previous analysis of the impact of the attacks, either there is a true obstacle or Attack PRA3 or Attack AO occurs.
    \end{proof}

We check whether $\mathcal{P}(\mathcal{O}^b_k(x_A, S_A)) \setminus \mathcal{U}_{Bk}$ is empty as follows. For each point $p \in \mathcal{P}(\mathcal{O}^b_k(x_A, S_A))$, we check to see whether $p$ satisfies $h_l^{Bk}(p) - ||a_{l}^{jk}||\zeta_{h} \leq 0, \forall l \in \{1,\ldots,n^h_{Bk}\}.$ If not, we put $p$ into $\mathcal{P}(\mathcal{O}^b_k(x_A, S_A)) \setminus \mathcal{U}_{Bk}$. We name this operation a scan match.
% We limit the scope of scan match of the observation to the set $\mathcal{P}(\mathcal{O}^b_k(x_A, S_A)) \setminus \mathcal{U}_B$, which is equal to the set of points that are observed by Agent A and are in the space that can be observed by Agent B. 
% We also limit the scope of scan match of the observation to the horizontal axes of two augmented observations $\Tilde{\mathcal{O}}_a^{\mathcal{I}}$ and $\Tilde{\mathcal{O}}_b^{\mathcal{I}}$. 

We design the decision tree as shown in Fig.~\ref{fig:FDII} to detect each attack. 
% At each time, Agent A collects the current scans of nearby agents $S_j, \forall j \in L, j \neq A$. The corresponding observations $\mathcal{O}(x_A, S_j),$ the observations of obstacles $\mathcal{O}^b_k(x_A, S_j),$ and the occupied areas $\mathcal{U}_j$ are calculated by Agent A. For each bounding box provided by the object detection algorithm, Agent A does a scan match based on the result of Lemma~\ref{lemma:scan-match}. 
Agent A iterates the bounding boxes to check whether an object is detected. The points in the bounding box belong to either the non-existing obstacle from Attack NEO or the true obstacle. After detecting and identifying the attack or authenticating that it is the true obstacle, Agent A removes the points from the observation, then move on to the next bounding box. After traversing all bounding boxes iteratively, Agent A checks whether there is still observation of the obstacle. If not, there is no more attack or obstacle. If there is the observation of the obstacle, there is an attack and a scan match will be done to decide the classification of the attack.

After detecting the scan mismatch and identifying the spoofed agent, the proposed FDII algorithm isolates the corrupted parts of the raw data and updates the unsafe region by extracting the intersection of the unions of the occupied areas  of the agents indexed in $L.$ 

The following theorem shows that the updated unsafe region $(\cup_{k = 1,\ldots,n^A_o} \mathcal{U}_{Ak}) \bigcap (\cup_{m = 1,\ldots,n^j_o} \mathcal{U}_{jm})$ contains $\mathcal{P}(P^{ob}_k), \forall k \in \{1,\ldots,n^A_o\}$. We note that Assumption~\ref{assmpt:no-obscure} is not required in Theorem~\ref{theorem:updated-unsafe-region}.
% The following theorem shows that the updated unsafe region $\bigcap_j (\cup_k \mathcal{U}_{jk})$ contains $\mathcal{P}(P^{ob}_k), \forall k \in \{1,\ldots,n^j_o\}, j \in L$. 
\begin{theorem}
\label{theorem:updated-unsafe-region}
Suppose we are given the occupied areas $\mathcal{U}_{Ak}$ of Agent A and $\mathcal{U}_{jk}$ of Agent $j$, $k \in \{1,\ldots,n^j_o\}$ in the area of $\mathcal{I}_A \cap \mathcal{I}_j$. If $\mathcal{P}(P^{ob}_k) \subseteq \mathcal{I}_A \cap \mathcal{I}_j$, then $\mathcal{P}(P^{ob}_k) \subseteq (\cup_{k = 1,\ldots,n^A_o} \mathcal{U}_{Ak}) \bigcap (\cup_{m = 1,\ldots,n^j_o} \mathcal{U}_{jm})$ for any of the attack types NEO, PRA, or AO. %If \textcolor{blue}{Attack NEO} occurs, then the existence of $P^{ob}_{1}$ will not enlarge $\bigcap_j (\cup_k \mathcal{U}_{jk}).$
\end{theorem}
% $\mathcal{P}(P^{ob}_k) \subseteq (\cup_k \mathcal{U}_{Ak}) \bigcap (\cup_k \mathcal{U}_{jk})$
% \begin{theorem}
% If the occupied areas $\mathcal{U}_{jk}$ of Agent $j$, $\forall j \in L, k \in \{1,\ldots,n^j_o\}$ in the area of $\cap_j \mathcal{I}_j$ are given, then $\mathcal{P}(P^{ob}_k) \subseteq \bigcap_j (\cup_k \mathcal{U}_{jk}).$ %If \textcolor{blue}{Attack NEO} occurs, then the existence of $P^{ob}_{1}$ will not enlarge $\bigcap_j (\cup_k \mathcal{U}_{jk}).$
% \end{theorem}

\begin{proof}
    % We have $\bigcup_{k}{\mathcal{U}_{Ak} \cap \mathcal{U}_{jk}} = (\cup_m \mathcal{U}_{Am}) \bigcap (\cup_k \mathcal{U}_{jk}), \forall m \in \{1,\ldots,n^A_o\}, \forall k \in \{1,\ldots,n^j_o\}.$
    First, we show that even if there is another obstacle $P^{ob}_{k^\prime}$ between $P^{ob}_k$ and Agent $j$, $\mathcal{P}(P^{ob}_k) \subseteq \cup_k \mathcal{U}_{jk}$ holds. %by relaxing Assumption~\ref{assmpt:no-obscure} in Lemma~\ref{lemma:occupied-area-covers-obstacle-projection}.

    For each obstacle $P^{ob}_k, \forall k \in \{1,\ldots,n^j_o\}$, we consider two cases, namely (i) partially or fully blocked case, in which there is another obstacle $P^{ob}_{k^\prime}$ between obstacle $P^{ob}_k$ and Agent $j$, and $P^{ob}_{k^\prime}$ partially or fully blocks $P^{ob}_k$ from Agent $j,$ and (ii) not blocked case, in which there is no $P^{ob}_{k^\prime}$ that blocks $P^{ob}_k$ from Agent $j.$

    \underline{\textit{Case (i)}}: In this case, there exists $k^{\prime} \in \{1,\ldots,n^j_o\}$ such that $\mathcal{P}(P^{ob}_k) \cap \mathcal{U}_{jk^{\prime}} \neq \emptyset.$ We divide $P^{ob}_k$ into two parts: $\mathcal{P}(P^{ob}_k) \cap \mathcal{U}_{jk^{\prime}}$ and $\mathcal{P}(P^{ob}_k) \setminus \mathcal{U}_{jk^{\prime}}.$ For $\mathcal{P}(P^{ob}_k) \cap \mathcal{U}_{jk^{\prime}}$, we have $$(\mathcal{P}(P^{ob}_k) \cap \mathcal{U}_{jk^{\prime}}) \subseteq \mathcal{U}_{jk^{\prime}} \subseteq \cup_k \mathcal{U}_{jk}.$$ 
    For $\mathcal{P}(P^{ob}_k) \setminus \mathcal{U}_{jk^{\prime}},$ since there is line-of-sight between $\mathcal{P}(P^{ob}_k) \setminus \mathcal{U}_{jk^{\prime}}$ and Agent $j$, Assumption~\ref{assmpt:no-obscure} holds. According to Lemma~\ref{lemma:occupied-area-covers-obstacle-projection}, we have $$(\mathcal{P}(P^{ob}_k) \setminus \mathcal{U}_{jk^{\prime}}) \subseteq \mathcal{U}_{jk} \subseteq \cup_k \mathcal{U}_{jk}.$$
    
    \underline{\textit{Case (ii)}}: In this case, $\mathcal{P}(P^{ob}_k) \cap \mathcal{U}_{jk^{\prime}} = \emptyset, \forall k^{\prime} \in \{1,\ldots,n^j_o\}.$ Since there is line-of-sight between $\mathcal{P}(P^{ob}_k)$ and Agent $j$, Assumption~\ref{assmpt:no-obscure} holds. According to Lemma~\ref{lemma:occupied-area-covers-obstacle-projection}, we have $\mathcal{P}(P^{ob}_k) \subseteq \mathcal{U}_{jk} \subseteq \cup_{m = 1,\ldots,n^j_o} \mathcal{U}_{jm}$. 
    
    We next show that even if one of the Attack NEO, PRA, or AO occurs, $\mathcal{P}(P^{ob}_k) \subseteq \cup_{m = 1,\ldots,n^j_o} \mathcal{U}_{jm}$. 

    For Attack NEO, $\mathcal{P}(P^{ob}_k) = \emptyset,$ then $\mathcal{P}(P^{ob}_k) \subseteq \cup_{m = 1,\ldots,n^j_o} \mathcal{U}_{jm}$. 
    
    For Attack PRA, we first consider the case where $e^\prime$ is not blocked by any obstacle. According to the threat model in Section~\ref{subsec:threat-model}, we have $e^{\prime}$ obscuring $P^{ob}_k$. Hence according to Case (i), we have $\mathcal{P}(P^{ob}_k) \subseteq \mathcal{U}_{jk} \subseteq \cup_{m = 1,\ldots,n^j_o} \mathcal{U}_{jm}$. We next consider the case when $e^\prime$ is partially or fully blocked by another obstacle $k^{\prime}$. According to Case (i), we have $\mathcal{P}(P^{ob}_k) \subseteq \mathcal{U}_{jk}\cup \mathcal{U}_{jk^{\prime}} \subseteq \cup_{m = 1,\ldots,n^j_o} \mathcal{U}_{jm}$.
    % For Attack PRA, we consider two cases: $e^\prime$ is not blocked by any Obstacle and $e^\prime$ is blocked by the Obstacle $k^\prime$. If $e^\prime$ is not blocked by any Obstacle, $\mathcal{P}(P^{ob}_k) \subseteq \mathcal{U}_{jk}$ according to the model for Attack PRA in Section~\ref{subsec:threat-model}, which means $P^{ob}_k$ is blocked by $e^\prime.$ According to the result of Case (i), we have $\mathcal{P}(P^{ob}_k) \subseteq \cup_{m = 1,\ldots,n^j_o} \mathcal{U}_{jm}$. If $e^\prime$ is blocked by the Obstacle $k^\prime$, then $\mathcal{P}(e^\prime) \cap \mathcal{U}_{jk^{\prime}} \neq \emptyset.$ 
    % For Attack PRA, $\mathcal{P}(P^{ob}_k) \subseteq \mathcal{U}_{jk}$ according to the model for Attack PRA in Section~\ref{subsec:threat-model}, where $\mathcal{U}_{jk}$ is the occupied area of $e^\prime$. According to the results of Case (i) and Case (ii), we have Thus $\mathcal{P}(P^{ob}_k) \subseteq \cup_{m = 1,\ldots,n^j_o} \mathcal{U}_{jm}$.
    % \underline{\textit{For Attack PRA}}, $\mathcal{P}(P^{ob}_k) \subseteq \mathcal{U}_{A{k^\prime}}, k^\prime \in \{1,\ldots,n^j_o\}.$ Thus $\mathcal{P}(P^{ob}_k) \subseteq \cup_k \mathcal{U}_{Ak}$ holds. 
    
    For Attack AO, %$\mathcal{P}(e^\prime) \subseteq \mathcal{P}(P^{op}_k) \subseteq \mathcal{U}_{jm}$. 
    obstacle $P^{op}_k$ could be treated as a true obstacle with observation $\mathcal{O}(x_j, e^\prime).$ %If $e^\prime$ is not obscured by other obstacles, we have $\mathcal{P}(P^{ob}_k) \subseteq \mathcal{U}_{jk} \subseteq \cup_{m = 1,\ldots,n^j_o} \mathcal{U}_{jm}$ according to Case (ii). Otherwise, 
    Hence, according to Case (i) and (ii), we have $\mathcal{P}(P^{ob}_k) \subseteq  \cup_{m = 1,\ldots,n^j_o} \mathcal{U}_{jm}$.
    % when there is line-of-sight between $P^{ob}_k$ and Agent $j,$ we have $\mathcal{P}(P^{ob}_k) \subseteq \mathcal{U}_{jk}$ based on Assumption~\ref{assmpt:no-obscure}. Thus $\mathcal{P}(P^{ob}_k) \subseteq \cup_{m = 1,\ldots,n^j_o} \mathcal{U}_{jm}$. 

    Since $\mathcal{P}(P^{ob}_k) \subseteq \mathcal{U}_{jk} \subseteq \cup_{m = 1,\ldots,n^j_o} \mathcal{U}_{jm}, j \in L$ holds for any of the attack types NEO, PRA, or AO, thus we have $\mathcal{P}(P^{ob}_k) \subseteq (\cup_{k = 1,\ldots,n^A_o} \mathcal{U}_{Ak}) \bigcap (\cup_{m = 1,\ldots,n^j_o} \mathcal{U}_{jm})$ for any of the attack types NEO, PRA, or AO.
\end{proof}

To calculate $(\cup_{k = 1,\ldots,n^A_o} \mathcal{U}_{Ak}) \bigcap (\cup_{m = 1,\ldots,n^j_o} \mathcal{U}_{jm}),$ which is equivalent to $\bigcup_{k^\prime = 1,\ldots,n^A_o} ( \mathcal{U}_{A{k^\prime}} \bigcap (\cup_{k = 1,\ldots,n^j_o} \mathcal{U}_{jk}))$, Agent A approximates the set $\mathcal{U}_{A{k^\prime}} \bigcap (\cup_{k = 1,\ldots,n^j_o} \mathcal{U}_{jk})$ by computing the convex hull of all scan points whose projections are contained in $\mathcal{U}_{A{k^\prime}} \bigcap (\cup_{k = 1,\ldots,n^j_o} \mathcal{U}_{jk}).$ Then Agent A computes a set of half-plane constraints $h^{jk^\prime}_l(x), l = \{1,\ldots,n^h_{jk^\prime}\}$ based on the vertices, where $h^{jk^\prime}_l(x)  - ||a_{l}^{jk}||\zeta_{h} < 0$ describes the corresponding half-plane and $n^h_{jk^\prime}$ is the number of half-planes forming the convex hull $\mathcal{U}_{jk^\prime}$. We define $\Bar{h}_{jk^\prime}(x) = \max_l h^{jk^\prime}_l(x)$.

\subsection{Safe Control}
In this subsection, we present the safe vehicle controller based on the unsafe region provided by the FDII. The  kinematic bicycle model~\cite{rajamani2011vehicle}  adopted in this paper is defined as 

\begin{equation}
    \label{eq:bicycle-model-continuous}
    \begin{bmatrix} 
    \Dot{\phi}\\
    \Dot{x}\\
    \Dot{y}\\
    \end{bmatrix}
    = \begin{bmatrix}
    \frac{v}{l}\sin{\psi}\\
    v\cos{(\phi + \psi)}\\
    v\sin{(\phi + \psi)}\\
    \end{bmatrix}
\end{equation}
where $\phi$ is the heading angle between the orientation and the x-axis, $(x, y)$ is the position of the reference point (at the middle of the front axle, between the front wheels), $v$ is the forward velocity at the reference point, $l$ is the wheelbase, and $\psi$ is the steering angle.

For implementation, we discretize Eq.~\ref{eq:bicycle-model-continuous} using forward differencing method with the sample time $dt$ to obtain
\begin{align}
    \label{eq:bicycle-model-discrete}
    \mathbf{x}[t+1] &= \begin{bmatrix} 
    \phi[t+1]\\
    x[t+1]\\
    y[t+1]\\
    \end{bmatrix}
    = f(\mathbf{x}[t], \mathbf{u}[t]) \nonumber\\
    &= \begin{bmatrix}
    \phi[t] + dt\frac{v[t]}{l}\sin{\psi[t]}\\
    x[t] + dtv[t]\cos{(\phi[t] + \psi[t])}\\
    y[t] + dtv[t]\sin{(\phi[t] + \psi[t])}\\
    \end{bmatrix}
\end{align}
where $\mathbf{u}[t] = (v[t], \psi[t])^\prime.$

In order to avoid collision between the vehicle and the unsafe region, we utilize Model Predictive Control with Discrete-Time Control Barrier Function (MPC-CBF)~\cite{zeng2021safety}. The controller solves the following finite-time optimization problem with prediction horizon $T$ at each time step
\begin{subequations}
\label{eq:MPC-DTCBF}
\begin{align}
    \label{eq:objective-equation}
    \min_{\mathbf{u}[t:t+T-1]} \ & (\mathbf{x}[t+T]-\mathbf{x}_r)^TF(\mathbf{x}[t+T]-\mathbf{x}_r) \\
    &+ \sum_{t^{\prime}=t}^{t+T-1} (\mathbf{x}[t^{\prime}]-\mathbf{x}_r)^TQ(\mathbf{x}[t^{\prime}]-\mathbf{x}_r) + \mathbf{u}[t^{\prime}]^TR\mathbf{u}[t^{\prime}] \nonumber\\
    \text{s.t.} \ & \Bar{h}_{jk^{\prime}}(\mathbf{x}[t+1])-\Bar{h}_{jk^{\prime}}(\mathbf{x}[t]) \geq -\gamma \Bar{h}_{jk^{\prime}}(\mathbf{x}[t]), \nonumber\\
    \label{eq:DT-CBF}
    & k^{\prime} = 1,\ldots,n^A_o \\
    % \text{s.t.} \ & h_i(\mathbf{x}[k+1])-h_i(\mathbf{x}[k]) \geq -\gamma h_i(\mathbf{x}[k]), \nonumber\\
    % \label{eq:DT-CBF}
    % & k = t,\cdots,t+T-1, \ i = 1,\cdots,n_h \\
    \label{eq:system-dynamics-equation}
    & \mathbf{x}[t^{\prime}+1] = f(\mathbf{x}[t^{\prime}], \mathbf{u}[t^{\prime}]), t^{\prime} = t,\cdots,t+T-1 \\
    \label{eq:admissible-equations-x}
    & \mathbf{x}[t^{\prime}] \in X, t^{\prime} = t,\cdots,t+T-1 \\
    \label{eq:admissible-equations-u}
    & \mathbf{u}[t^{\prime}] \in U, t^{\prime} = t,\cdots,t+T-1 
\end{align}
\end{subequations}
where Eq.~\eqref{eq:objective-equation} means that we would like the vehicle to converge to the midline of the lane ($x_r$) with minimal control effect, Eq.~\eqref{eq:DT-CBF} is the DT-CBF constraints ensuring that the vehicle avoids the unsafe regions provided by FDII, $n^j_o$ is the number of the occupied areas of Agent $j$, Eq.~\eqref{eq:system-dynamics-equation} is the system dynamics as shown in Eq.~\eqref{eq:bicycle-model-discrete}, Eq.~\eqref{eq:admissible-equations-x} is the admissible set of system state $\mathbf{x}$ (eg. stationary obstacles shown in the default map), and Eq.~\eqref{eq:admissible-equations-u} is the admissible set of control input $\mathbf{u}$ (eg. the upper bounds and lower bounds of the control inputs). 

The optimal solution of Eq.~\eqref{eq:MPC-DTCBF} is a sequence of control inputs $\mathbf{u}^{\ast}[t], \ldots, \mathbf{u}^{\ast}[t+T-1]$. The first element $\mathbf{u}[t] = \mathbf{u}^{\ast}[t]$ will be executed. Since $\mathbf{u}[t] = \mathbf{u}^{\ast}[t]$ satisfies Eq.~\eqref{eq:DT-CBF}, the vehicle will not collide with the obstacle at time step $t.$ Then at time step $t+1,$ Eq.~\eqref{eq:MPC-DTCBF} will be solved based on the new state $x[t+1]$. If FDII provides updated unsafe regions, they will be incorporated in Eq.~\eqref{eq:DT-CBF}. This receding horizon control strategy guarantees collision avoidance between the vehicle and the obstacles.
% We next design a safe control methodology to ensure safety of the controlled agent with desired probability. We formulate it as a control of a hybrid system. 

% In section \ref{sec:FDII}, we detect and estimate the obstacle to acquire unsafe region $h_1(x)$. Hence, we have the safe set as $\mathcal{C}_1:=\{x\mid h_1(x)\geq0\}$. 
\section{Case Study}
\label{sec:case-study}

In this section, we evaluate our approach in CARLA \cite{dosovitskiy2017carla} simulation environment. We first evaluate our proposed approach to FDII and obstacle detection under attacks. 
We then show that our safe controller ensures that the CARLA vehicle avoids obstacles using the proposed control strategy. 

\subsection{Augmented LiDAR FDII}
\label{subsec:sim_FDII}
We first introduce the simulation settings. We initialize two vehicles, denoted as Agents A and B, at locations $(-54.34, 137.05)$ and $(-34.34, 137.05)$, respectively. We evaluate our Augmented LiDAR FDII by simulating four scenarios: attack-free, attack NEO, PRA2 and PRA3. 

\begin{figure*}[!htbp]
\centering
\begin{subfigure}{.3\textwidth}
    \centering
    \includegraphics[width = \textwidth]{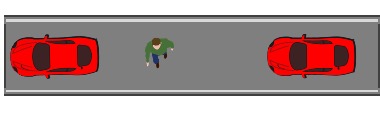}
    \subcaption{Attack-free: a pedestrian is set between two agents without attack.}
    \label{fig:set_Attack-free}
\end{subfigure}%
\hfill
\begin{subfigure}{.3\textwidth}
    \centering
    \includegraphics[width = \textwidth]{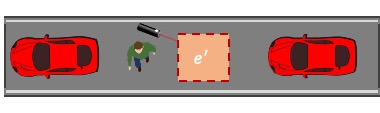}
    \subcaption{Attack PRA2: The attack signal is set between Agent A and the pedestrian.}
    \label{fig:set_Attack-PRA2}
\end{subfigure}%
\hfill
\begin{subfigure}{.3\textwidth}
    \centering
    \includegraphics[width = \textwidth]{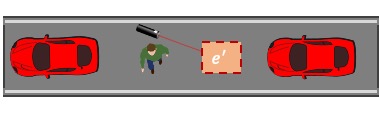}
    \subcaption{Attack PRA3: The attack signal affects a relatively smaller area.}
    \label{fig:set_Attack-PRA3}
\end{subfigure}%
\vfill

\begin{subfigure}{.3\textwidth}
    \centering
    \includegraphics[width = \textwidth]{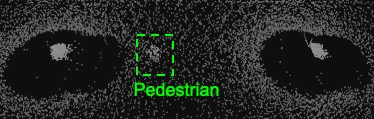}
    \subcaption{Annotated pedestrian can be detected by both agents with $\mathcal{U}_A\neq\emptyset$ and $\mathcal{U}_B\neq\emptyset$. }
    \label{fig:pcd_Attack-free}
\end{subfigure}
\hfill
\begin{subfigure}{.3\textwidth}
% \title{Testing Episodes}
  \centering
  \includegraphics[width = \textwidth]{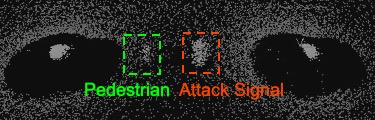}
  \subcaption{Annotated attack signal is captured by Agent A with $\mathcal{U}_A\neq\emptyset$ and $\mathcal{U}_B\neq\emptyset$. }
  \label{fig:pcd_Attack-PRA2}
\end{subfigure}%
\hfill
\begin{subfigure}{.3\textwidth}
% \title{Testing Episodes}
  \centering
  \includegraphics[width = \textwidth]{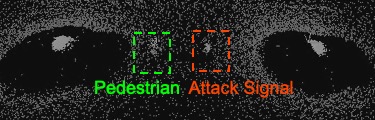}
  \subcaption{Annotated attack signal is captured by Agent A with $\mathcal{U}_A\neq\emptyset$ and $\mathcal{U}_B\neq\emptyset$. }
  \label{fig:pcd_Attack-PRA3}
\end{subfigure}%
\vfill

\begin{subfigure}{.3\textwidth}
    \centering
    \includegraphics[width = \textwidth]{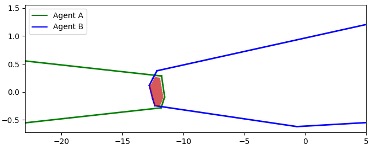}
    \subcaption{The FDII module detects no attack. The intersection of the occupied area $\mathcal{U}_A\cap\mathcal{U}_B$ identified as an unsafe region. }
    \label{fig:ocp_Attack-free}
\end{subfigure}
\hfill
\begin{subfigure}{.3\textwidth}
% \title{Testing Episodes}
  \centering
  \includegraphics[width = \textwidth]{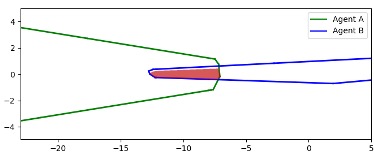}
  \subcaption{The FDII module detects PRA2 attack. The intersection of the occupied area $\mathcal{U}_A\cap\mathcal{U}_B$ identified as an unsafe region. }
  \label{fig:ocp_Attack-RPA2}
\end{subfigure}%
\hfill
\begin{subfigure}{.3\textwidth}
% \title{Testing Episodes}
  \centering
  \includegraphics[width = \textwidth]{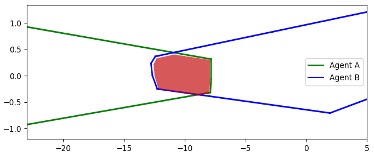}
  \subcaption{The FDII module detects PRA3 attack. The intersection of the occupied area $\mathcal{U}_A\cap\mathcal{U}_B$ identified as an unsafe region. }
  \label{fig:ocp_Attack-RPA3}
\end{subfigure}%
\caption{Augmented LiDAR FDII simulation settings and results: We demonstrate settings in the first row, the corresponding point cloud of joint perception of two agents and the candidate unsafe region in the second and third row, respectively. We list attack-free, PRA2 and PRA3 in the three columns from left to right, respectively.}
\label{fig:FDII_sim_res}
\end{figure*}

We simulate the attack-free scenario as a reference group. As shown in Fig. \ref{fig:set_Attack-free}, a pedestrian spawns at $(-42.34, 137.05)$, $12$ meters in front of Agent A. 

\begin{figure}[!htbp]
\centering
    \begin{subfigure}{.3\textwidth}
        \centering
        \includegraphics[width = \textwidth]{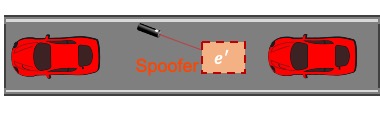}
        \subcaption{Attack NEO: An obstacle falsified by spoofer is set in front of Agent A.}
        \label{fig:set_Attack-NEO}
    \end{subfigure}%
    \vfill
    \begin{subfigure}{.3\textwidth}
    % \title{Testing Episodes}
      \centering
      \includegraphics[width = \textwidth]{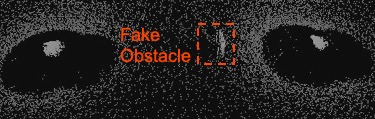}
      \subcaption{Fake obstacle can be detected only by Agent A with $\mathcal{U}_A\neq\emptyset$ and $\mathcal{U}_B=\emptyset$.}
      \label{fig:pcd_Attack-NEO}
    \end{subfigure}%
    \vfill
    \begin{subfigure}{.3\textwidth}
    % \title{Testing Episodes}
      \centering
      \includegraphics[width = \textwidth]{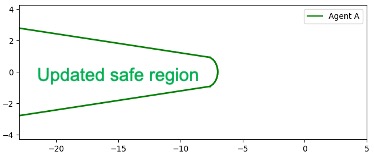}
      \subcaption{The FDII module detects NEO attack. The intersection of the occupied area $\mathcal{U}_A\cap\mathcal{U}_B=\emptyset$, denoting there is no unsafe region.}
      \label{fig:ocp_Attack-NEO}
    \end{subfigure}%
    \caption{FDII simulation settings and results of attack-NEO case}
\end{figure}

In Attack NEO, the attacker injects false data to create a non-existing obstacle. As shown in Fig. \ref{fig:set_Attack-NEO}, the false data is injected into Agent A's LiDAR observation to falsify a cylinder $8$ meters in front of agent A with radius $1$ meter.

The PRA attacker spoofs LiDAR detection modules to hide obstacles in relay attacked area by injecting false data between victim agent and obstacles. We simulate this attack by replacing true measurement with Gaussian noise. In attack PRA2, we use the same basic setting as the aforementioned attack-free case. In addition, we let the attacker inject false data into Agent A's LiDAR observation to hide the pedestrian in the relay attacked area as shown in Fig. \ref{fig:set_Attack-PRA2}. The injected false data, located $8$ meters ahead of Agent A, is Gaussian noise distributed in a cylinder area with $1.3$ meters radius. 

In the previous PRA2 case, the cylinder area can be partially seen by Agent B. We further validate our approach on a scenario where the cylinder area is blocked by pedestrian. In Attack PRA3 case, we set the cylinder area with $0.4$ meter radius, shown in Fig. \ref{fig:set_Attack-PRA3}. 

% In the first scenario, we set two vehicles driving in the same direction. The distance between them is 20 meters. A pedestrian and injected false data are set 12 and 8 meters in front of the rear vehicle, respectively. The false data is set in a cylinder whose radius is 0.4 meters in situation b3/c and 1.3 meters in situation b1/b2. The location of points in this cylinder is Gaussian distribution. We use the Gaussian distribution to simulate the physical removal attack.

% In the second scenario, two vehicles drive in the same direction, and the distance between them is set as 20 meters. The false data is set 8 meters in front of the rear vehicle. The false data is introduced as the surface of a cylinder, whose radius is 1 meter, facing the rear car.

Finally, we present the simulation results and analyze them by comparing with the reference attack-free scenario. We list the corresponding point cloud of joint perception of two agents and the intersection of the occupied area in the second and third row of Fig. \ref{fig:FDII_sim_res}, respectively. 

In the attack-free case, agents exchange point cloud data of the intersected area.  Both agents can detect the obstacle and generate bounding boxes to contain those points. Agents identify occupied area and generate non-empty sets $\mathcal{U}_A$ and $\mathcal{U}_B$ with the contained points. The proposed FDII module takes these information and detects that there is no attack. By overlapping $\mathcal{U}_A$ and $\mathcal{U}_B$ according to agents' location, FDII further identifies the intersection shown in Fig. \ref{fig:ocp_Attack-free} as the candidate unsafe region. 

In the attack NEO case, a fake obstacle is detected by Agent A annotated in Fig. \ref{fig:pcd_Attack-NEO}, which create an erroneous unsafe region. Due to the aforementioned limitation of relay attack, this fake obstacle can only be seen by Agent A. Therefore with the observation provided by Agent B, there is no occupied area identified with the points contained, i.e., $\mathcal{U}_B=\emptyset$. The proposed FDII module detects attack NEO and the obstacle is non-existing. Hence, the unnecessary unsafe region is removed, shown in Fig. \ref{fig:ocp_Attack-NEO}.

In attack PRA2 case, both attack signal and the pedestrian can be captured by Agent A and B respectively. With point cloud from Agent B, Agent A can observe both obstacles annotated in Fig. \ref{fig:pcd_Attack-PRA2} and generate their corresponding occupied area with $\mathcal{U}_A\neq\emptyset$ and $\mathcal{U}_B\neq\emptyset$. Since some areas affected by injected false data can be observed by Agent B, we have the $\mathcal{P}(\mathcal{O}^b_k(x_A, S_A)) \nsubseteq \mathcal{U}_B$ and $\mathcal{P}(\mathcal{O}^b_k(x_A, S_A)) \cap \mathcal{U}_B \neq \emptyset$. The proposed FDII algorithm detects attack PRA2. By overlapping $\mathcal{U}_A$ and $\mathcal{U}_B$ according to agents' location, FDII further identifies the intersection shown in Fig. \ref{fig:ocp_Attack-RPA2} as the candidate unsafe region.  

In attack PRA3 case, the false data affected area is relatively small as shown in Fig. \ref{fig:pcd_Attack-PRA3}, and hence the area is fully blocked by the pedestrian from Agent B's perspective. In this case, we have the $\mathcal{P}(\mathcal{O}^b_k(x_A, S_A)) \subseteq \mathcal{U}_B$ and $\mathcal{P}(\mathcal{O}^b_k(x_A, S_A)) \cap \mathcal{U}_B \neq \emptyset$. The proposed FDII algorithm detects attack PRA3. By overlapping $\mathcal{U}_A$ and $\mathcal{U}_B$ according to agents' location, FDII further identifies the intersection shown in Fig. \ref{fig:ocp_Attack-RPA3} as the candidate unsafe region.  
% In the simulated attack PRA3 scenario, an unsafe region generates in the augmented point cloud shown in Fig. \ref{fig:pcd_Attack-PRA3} between the pedestrian and the attack signal. The FDII algorithm outputs $4$, denoting the case is under relay attack, and the hidden obstacle is fully covered because of the false data.

\subsection{Safe Control}
\label{subsec:sim_control}
In this case study, we show that our MPC controller ensures the vehicle to be safe. We consider CARLA vehicle Model-3 as our control object and use \eqref{eq:bicycle-model-discrete} as the simplified vehicle model with $l=4$ and $dt=0.03$. We further take standard feedback linearization approach to acquire linear model as 
\begin{align}
    \label{eq:fblinear_model}
    \begin{bmatrix} 
    x\\
    y\\
    v_x\\
    v_y\\
    \end{bmatrix}_{k+1}
    = & \left[ \begin{array}{cccc}
    1 & 0 & 0.03 & 0 \\
    0 & 1 & 0 & 0.03 \\
    0 & 0 & 1 & 0 \\ 
    0 & 0 & 0 & 1 \end{array} \right]
    \begin{bmatrix} 
    x\\
    y\\
    v_x\\
    v_y\\
    \end{bmatrix}_k \\
    & + \left[ \begin{array}{cccc}
    0.0045 & 0 \\
    0 & 0.0045 \\
    1 & 0 \\ 
    0 & 1 \end{array} \right]
    \begin{bmatrix} 
    \Delta v_x\\
    \Delta v_y\\
    \end{bmatrix}_k, 
\end{align}
in which $v_x$, $v_y$ are velocity component of x-axis and y-axis, respectively, control input $u=[\Delta v_x, \Delta v_y]^T$ representing the corresponding changes. 

We define an MPC controller according to \eqref{eq:MPC-DTCBF} with $F$, $Q$, and $R$ set to be identical matrices. We realize our controller with an open-source Python library do-mpc \cite{lucia2017rapid}, which calls CasADi \cite{andersson2019casadi} and IPOPT \cite{wachter2006implementation} for nonlinear programming. 

We next present the setting of the case. The vehicle is asked to perform reach-and-avoid task starting from location $(-14.34, 137.05)$ to $(-5.00, 135.25)$ without entering the unsafe region. We set the initial state to be $[-14.34, 137.05, 0, 0]^T$ and initial guess to be $[1,0]^T$. Given the unsafe region detected by FDII module, controller restores the constraints on position $\boldsymbol{s} = [x,y]^T$ as
\begin{align*}
    h(\boldsymbol{s})_1 &= [-0.35, 0.94] \boldsymbol{s} - 132.74 \\
    h(\boldsymbol{s})_2 &= [-0.17, -0.99] \boldsymbol{s} + 132.5 \\
    &\ldots \\
    h(\boldsymbol{s})_{14} &= [0.12, -0.99] \boldsymbol{s} - 134.62. 
\end{align*}

\begin{figure}
    \centering
    \includegraphics[width = 0.4\textwidth]{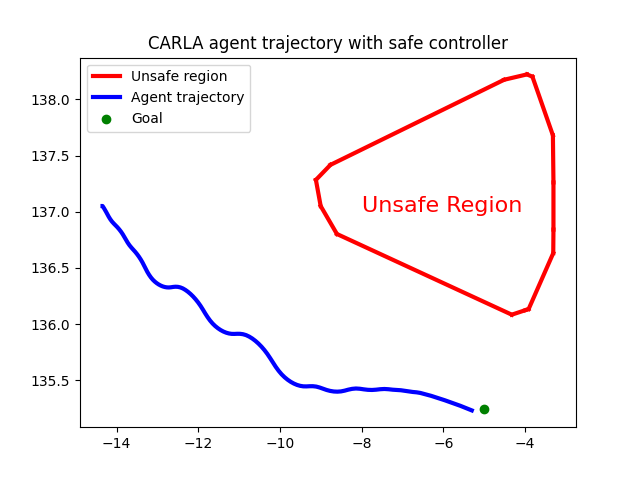}
    \caption{MPC drove CARLA vehicle from start $(-14.34, 137.05)$ to goal $(-5.00, 135.25)$. The agent managed to avoid detected unsafe region while tracking the given reference point. }
    \label{fig:MPC}
\end{figure}

Finally, we present the trajectory of the CARLA vehicle controlled by MPC in an urban street. As shown in Fig. \ref{fig:MPC}, the vehicle drives from the location $(-14.34, 137.05)$ to $(-5.00, 135.25)$ without entering the unsafe region. 

\section{Conclusion}
\label{sec:conclusion}
This paper presented an approach for leveraging sensor data from neighboring vehicles to detect LiDAR spoofing attacks on autonomous vehicles. In our approach, vehicles exchange LiDAR scan data and identify spoofing attacks by checking for disparities between the detected obstacles under each scan. We further develop a decision tree to differentiate between non-existing obstacle, physical removal, and adversarial object attacks. We then construct an estimate of the unsafe region based on the joint scan data, and propose a control policy that avoids the unsafe region. We validated our framework using the CARLA simulation platform and showed that it can detect and identify LiDAR attacks as well as guarantee safe driving.

% \section*{Acknowledgment}



%\bibliographystyle{IEEEtranS}

\bibliographystyle{IEEEtranS}
% \bibliography{IEEEabrv,ref}
\bibliography{ref}

\end{document}